\documentclass[a4paper,11pt]{article}
\usepackage{jheppub} 
\usepackage{mymacros}
\definecolor{change}{rgb}{0.0, 0.0, 0.0} 

\arxivnumber{2401.01930} 

\title{\boldmath Entanglement and confinement in lattice gauge theory tensor networks}

\author[1]{Johannes Knaute\note{Corresponding author.},}
\author{Matan Feuerstein}
\author{and Erez Zohar}
\affiliation{Racah Institute of Physics, The Hebrew University of Jerusalem,\\
Jerusalem 91904, Givat Ram, Israel}

\emailAdd{johannes.knaute@mail.huji.ac.il}

\abstract{We develop a transfer operator approach for the calculation of R\'enyi entanglement entropies in arbitrary (i.e.\ Abelian and non-Abelian) pure lattice gauge theory projected entangled pair states in 2+1 dimensions. It is explicitly shown how the long-range behavior of these quantities gives rise to an entanglement area law in both the thermodynamic limit and in the continuum. We numerically demonstrate the applicability of our method to the $\mathds{Z}_2$ lattice gauge theory and relate some entanglement properties to the confinement-deconfinement transition therein. We provide evidence that R\'enyi entanglement entropies in certain cases do not provide a complete probe of (de)confinement properties compared to Wilson loop expectation values as other genuine (nonlocal) observables.
}

\begin{document}
\maketitle
\flushbottom

\section{Introduction and motivation}
\label{sec:intro}

Characterization of entanglement properties in quantum field and gauge theories is of upmost importance for the understanding of physical systems in condensed matter, particle and gravitational physics \cite{Witten:2018zxz,Chen:2021lnq,Catterall:2022wjq}. As such, it allows to gain insights into important properties of emergent phenomena. In this article, we are particularly interested in the interplay of entanglement and confinement. The latter is a nonperturbative phenomenon, which describes the binding, i.e.\ confinement, of static charges. It appears prominently in quantum chromodynamics (QCD) as the theory of strong interactions in the standard model of particle physics, for which it manifests itself through the existence of color-neutral and strongly interacting quark bound states (e.g.\ hadrons and mesons) at low energies \cite{Wilson:1974sk}. On the other hand, at high energies, a deconfined quark-gluon plasma exists as a result of asymptotic freedom, for which a perturbative treatment can be possible \cite{Gross:1973ju,Busza:2018rrf,Berges:2020fwq}. Examples of confinement occur also in lower-dimensional quantum many-body (QMB) systems and quantum field theories (QFTs) \cite{McCoy:1978ta,Zamolodchikov:1989fp,Fonseca:2006au,Kormos2017}. They provide a way to study many long-standing open questions in this context using novel tools from a quantum information perspective.

One main motivation for our work comes from holography and the AdS/CFT correspondence \cite{Maldacena:1997re,Witten:1998qj,Gubser:1998bc}. The seminal work \cite{Ryu:2006bv,Ryu:2006ef} provided a holographic prescription for entanglement entropy, as the most prominent and rigorous entanglement measure of pure states, by showing that it can be calculated as the minimal surface area in the bulk belonging to a chosen boundary region. When studying this quantity, the authors of \cite{Klebanov:2007ws} proposed that entanglement (entropy) can serve as a probe of the confinement-deconfinement transition in gravitational duals of large-$N_c$ gauge theories. This idea was further studied and confirmed in many holographic models for QCD and beyond, see e.g.~\cite{Lewkowycz:2012mw,Kim:2013ysa,Kol:2014nqa,Ghodrati:2015rta,Zhang:2016rcm,Knaute:2017lll,Ali-Akbari:2017vtb,Dudal:2018ztm,Li:2020pgn,Arefeva:2020uec}. 
Following this line of research, several recent works unveiled the importance of entanglement properties also directly in processes relevant for particle and nuclear physics \cite{Berges:2017hne,Feal:2018ptp,Bellwied:2018gck,Liu:2022hto,Liu:2022qqf,Ehlers:2022oal,Asadi:2022vbl,Asadi:2023bat,Almeida:2023tak,Gu:2023aoc,Dosch:2023bxj,Shivashankara:2023koj}. 

In this work, we are using tensor network methods to analyze entanglement properties in (2+1)-dimensional lattice gauge theories (LGTs).\,\footnote{Note that when studying entanglement measures, we are considering time-independent setups. We are therefore commonly referring to 2$D$ LGTs throughout this paper.} Tensor networks and tensor network states are a concept and algorithmic tool originating in quantum information science. They are formulated in a Hamiltonian framework and provide efficient ans\"atze for wavefunctions of QMB states with a polynomial number of parameters. (For broad reviews on that topic see \cite{Verstraete08,SCHOLLWOCK201196,Orus:2013kga,Silvi:2017srb,Cirac:2020obd,Okunishi2022review,Banuls2022review}.) In contrast to Monte-Carlo simulations in Euclidean spacetime, tensor networks are sign-problem-free and usable for real-time simulations. They also allow to implement gauge symmetries as the necessary basis for the study of gauge theories in particle and condensed matter physics \cite{Banuls:2019rao,Banuls:2019bmf}. In one spatial dimension, matrix product states (MPS) have been successfully used to study a plethora of physical effects in spin chain models, QFTs and LGTs, see e.g.\ the recent reviews \cite{Banuls:2019rao,Banuls:2019bmf,Cirac:2020obd,Okunishi2022review,Banuls2022review} and references therein.
Specifically, they were also employed to study the effect of confinement on static and dynamical entanglement properties \cite{Kormos2017,Magnifico:2019kyj,Chanda:2019fiu,Castro-Alvaredo:2020mzq,Rigobello:2021fxw,Halimeh:2022pkw,Banuls:2022iwk,Knaute:2022ljb,Kebric:2023nnd,Osborne:2023zjn,Khor:2023xar}.
In two dimensions, projected entangled pair states (PEPS) are the natural generalization of MPS \cite{Verstraete08,Orus:2013kga,Cirac:2020obd}. Several gauging mechanisms have been proposed for them \cite{Haegeman:2014maa,Zohar:2015jnb}, which allow to study LGTs in (2+1)$D$. For example, the works \cite{Zapp:2017fcr,Robaina:2020aqh} analyzed infinite systems using a translationally invariant (uniform) PEPS ansatz. The authors of \cite{Zohar:2015eda,Zohar:2016wcf,Emonts:2020drm,Emonts:2022yom} combined the gauging principle with a Gaussian fermionic PEPS, which can be efficiently contracted using Monte-Carlo methods \cite{Zohar:2017yxl}. Beyond PEPS, also other Hamiltonian tensor network types \cite{Tagliacozzo:2010vk,Tagliacozzo:2014bta,Felser:2019xyv,Magnifico:2020bqt,Montangero:2021puw,Cataldi:2023xki} and non-Hamiltonian partition function approaches with tensor networks \cite{Meurice:2020pxc} have been employed for the study of higher-dimensional LGTs.

When studying entanglement properties for a bipartition of a physical system into a subsystem $A$ and its complement $B$ via entropic quantities, the reduced density matrix $\rho_A \equiv \Tr_B \rho$ is calculated by taking a partial trace of the full density matrix $\rho$ over the Hilbert space of the complement region. This procedure assumes the separability of the underlying full Hilbert space into a direct product form, i.e.\ $\mathcal H_A \otimes \mathcal H_B$. In LGTs, entanglement considerations are complicated by the fact that the space of physical states does not admit such a direct product structure: As a consequence of the \textit{Gauss law}, which follows from demanding local symmetry constraints, there exist nonlocal degrees of freedom hindering this decomposition. Several approaches have been developed to overcome this obstacle and analyze entanglement properties in LGTs \cite{Velytsky:2008rs,Buividovich:2008kq,Buividovich:2008gq,Donnelly:2011hn,Casini:2013rba,Radicevic:2014kqa,Ghosh:2015iwa,Aoki:2015bsa,Soni:2015yga,VanAcoleyen:2015ccp,Itou:2015cyu,Casini:2015dsg,Soni:2016ogt,Anber:2018ohz,Rabenstein:2018bri,Panizza:2022gvd,Xu:2023zsz}.\footnote{Interestingly, the work \cite{Rabenstein:2018bri} confirmed the holographic prediction for the connection between entanglement and confinement for (3+1)$D$ SU($N_c$) Yang--Mills theory.} In the present article we work within an extended Hilbert space approach \cite{Buividovich:2008gq,Donnelly:2011hn,Ghosh:2015iwa,Aoki:2015bsa,Soni:2015yga}, which is obtained by taking the tensor product of all gauge field Hilbert spaces on the links of the lattice. A partial trace over the complement links is then well-defined. Alternatively, one can consider the entanglement for superselection sectors \cite{Casini:2013rba,Radicevic:2014kqa}, which is underlying a symmetry-resolved entanglement approach \cite{Goldstein:2017bua,Feldman:2022hrf} that is developed independently in the companion paper \cite{resolved_project} for LGTs. 

In pure LGTs, Wilson loops -- operators creating closed flux lines -- are important nonlocal, gauge-invariant observables. The decay of their expectation value allows to probe confinement properties: While a Wilson loop area law is present in the confined phase, a perimeter law scaling implies the existence of deconfined static charges. In the recent work \cite{Zohar:2021wqy} it was shown how the gauged PEPS setup introduced above can be employed to study Wilson loops and hence (de)confinement properties for arbitrary (i.e.\ Abelian and non-Abelian) pure LGTs. This construction was based on a transfer operator approach. Specifically, it was deduced how local properties of transfer operators, which are constructed out of the tensors obeying the gauge symmetry, dictate the long-range behavior of the Wilson loop. Here, we are following a parallel route to show how (normalized) R\'enyi entanglement entropies can be constructed for the two-dimensional LGT PEPS using transfer operators. We explicitly show how an entanglement area law \cite{Eisert:2008ur} emerges in the thermodynamic limit from our general result and persists in the continuum.\footnote{Note that an entanglement area law, in contrast to the Wilson loop area law, refers to the scaling w.r.t.\ the boundary area of a chosen subregion. Hence, it implies a perimeter law scaling for two-dimensional regions. We will differentiate these cases throughout this paper by explicitly referring to the Wilson loop or entanglement area law.} 
Based on these general results, we will argue that R\'enyi entanglement entropies are not an equally rigorous measure of confinement or deconfinement as the Wilson loop. However, using numerical techniques, we analyze how confinement properties do leave imprints on entanglement measures for the $\mathds{Z}_2$ LGT as a specific example. 

We finally would like to add that our explorations are also motivated by recent advances of quantum simulation experiments \cite{Cirac2012,Alexeev:2020xrq,Fraxanet:2022wgf}, in which confinement properties such as meson physics have already been explored \cite{Tan:2019kya,Mildenberger:2022jqr,Lamb:2023eyr,Zhang:2023hzr} or are at reach of current technologies \cite{Domanti:2023qht}. At the same time, quantum simulations also allow to measure and explore entanglement properties of quantum many-body systems and LGTs \cite{Islam:2015mom,Dalmonte:2017bzm,Brydges_2019,Kokail:2020opl,Kokail:2021ayb,Mueller:2022xbg}. In this line of research, we are addressing in this paper the timely topic of the interplay of both of these phenomena for two-dimensional LGTs, and want to provide further stimulus for its experimental exploration. 

This paper is organized as follows. In section~\ref{sec:TNS} we review the underlying gauged PEPS ansatz and elaborate on the construction of transfer operators. In section~\ref{sec:area} we develop the transfer operator approach for the calculation of R\'enyi entanglement entropies. This idea is at first described in 1$D$ for MPS and then generalized to 2$D$ for PEPS. We subsequently calculate R\'enyi entanglement entropies by contracting transfer operator plaquettes within a 2$D$ lattice and derive their long-range properties. In section~\ref{sec:Z2} this formalism is numerically applied to the $\mathds{Z}_2$ LGT and contrasted with confinement properties therein. Concluding discussions and an outlook are given in section~\ref{sec:summary}. The appendix~\ref{app:proof} contains mathematical derivations of properties of a dominant transfer matrix eigenvalue.

\section{Tensor network construction}
\label{sec:TNS}

In this section we review important concepts of LGTs and the gauged PEPS ansatz. We then outline the construction of transfer operators in this framework. We intend to focus on the most crucial properties, which are relevant for the following sections in this paper. For further mathematical background and a comprehensive discussion of LGTs themselves, we refer to \cite{Zohar:2021wqy}, whose notation we will mainly follow in this section.

\subsection{LGTs and gauge invariant PEPS}

\begin{figure}[t]
    \centering
   \includegraphics[width=0.75\textwidth]{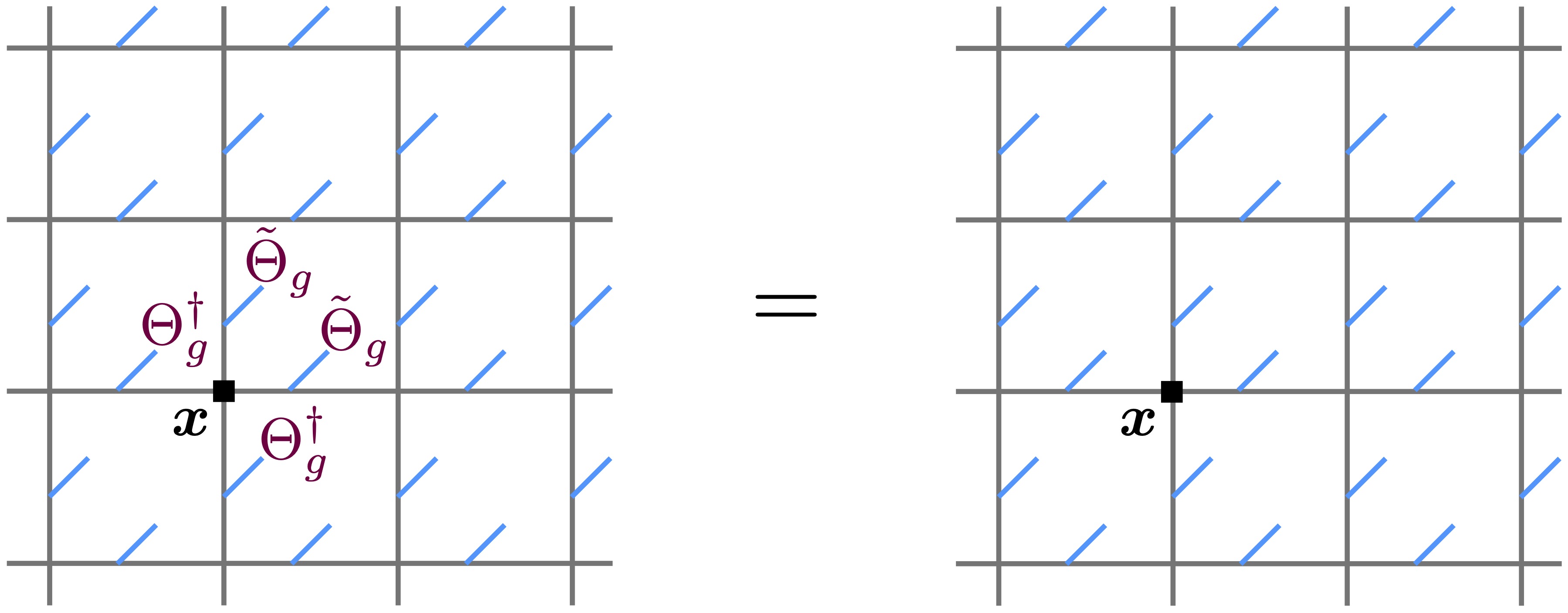}
    \caption{For a pure (2+1)$D$ LGT in the Hamiltonian formalism, the gauge fields are placed on the links of a spatial lattice (indicated by the blue lines). When acting with the gauge transformation \eqref{eq:gauge_trafo_general} on all four links around any lattice site $\bm{x}$, the represented quantum state remains invariant, cf.~eq.\,\eqref{eq:gauge_inv_state}.}
    \label{fig:lattice_inv}
\end{figure}

LGTs describe the interaction of matter particles through the mediation by gauge fields. In this paper we are studying (2+1)$D$ LGTs in a Hamiltonian formalism \cite{Kogut:1974ag}, for which the spatial coordinates are discretized, while time is kept continuous. We restrict ourselves to the pure gauge case, in which the gauge fields are placed on the links of a 2$D$ lattice, while fermionic matter fields, which would reside on the lattice sites, are absent; cf.\ Fig.~\ref{fig:lattice_inv}. 

Let us denote the lattice sites as $\bm{x} \in \mathds{Z}^2$, the directional unit vectors as $\hat{\bm{e}}_i$, and the emanating links as $\mathtt{l} \equiv (\bm{x},i)$, where $i=1,2$ labels the horizontal and vertical direction, respectively. By definition, LGTs are invariant under a local, i.e.\ spacetime-dependent symmetry. We introduce a gauge group $\mathcal G$ and denote its group elements as $g \in \mathcal G$. Each link $\mathtt{l}$ hosts a gauge field Hilbert space $\mathcal{H}_\mathtt{l}$, which can be either spanned by the gauge group element states $\ket{g}$, or, conventionally, by dual representation basis states $\ket{jmn}$. For these, $j$ labels an irreducible representation (irrep) of $\mathcal G$, and $m,n$ label identifiers within each.

Let us consider the \textit{gauge transformation}
\begin{equation} \label{eq:gauge_trafo_general}
    \hat\Theta_g(\bm{x}) = \tilde\Theta_g({\bm{x},1}) \, \tilde\Theta_g({\bm{x},2}) \, \Theta_g^\dagger({\bm{x}-\hat{\bm{e}}_1,1}) \, \Theta_g^\dagger({\bm{x}-\hat{\bm{e}}_2,2})
\end{equation}
which acts on the gauge fields at all four links around a lattice site $\bm{x}$ for a given group element. Here, $\Theta_g$ and $\tilde\Theta_g$ are unitary operators, which respectively realize right and left group multiplications on representation basis states, i.e.\ $\Theta_g\ket{jmn} = \ket{jmn'}D^j_{nn'}(g)$ and $\tilde\Theta_g\ket{jmn} = D^j_{mm'}(g)\ket{jm'n}$, where $D^j(g)$ is a unitary matrix representing the group element $g$ for the irrep $j$. A quantum state $\ket{\Psi}$ within the LGT (without static charges) is physical, if it satisfies the gauge invariance condition
\begin{equation} \label{eq:gauge_inv_state}
    \hat\Theta_g(\bm{x}) \ket{\Psi} = \ket{\Psi} \qquad \forall \bm{x}\in\mathds{Z}^2, g\in\mathcal{G} .
\end{equation}
See Fig.~\ref{fig:lattice_inv} for a pictorial representation. Analogously, gauge invariant operators $O$ satisfy
\begin{equation}
    \hat\Theta_g(\bm{x}) O \hat\Theta_g^\dagger(\bm{x}) = O \qquad \forall \bm{x}\in\mathds{Z}^2, g\in\mathcal{G} .
\end{equation}
For a compact Lie group $\mathcal G$, for which the operators can be written as $\Theta_g = \exp(i\phi_a(g)R_a)$ and $\tilde\Theta_g = \exp(i\phi_a(g)L_a)$ in terms of their right ($R_a$) and left transformation generators ($L_a$) and group element dependent parameters $\phi_a(g)$, the gauge transformation \eqref{eq:gauge_trafo_general} can be reexpressed using the \textit{Gauss law operator}
\begin{equation}
    G_a(\bm{x}) = L_a({\bm{x},1}) + L_a({\bm{x},2}) - R_a({\bm{x}-\hat{\bm{e}}_1,1}) - R_a({\bm{x}-\hat{\bm{e}}_2,2}) .
\end{equation}
The gauge invariance condition \eqref{eq:gauge_inv_state} is then reformulated in terms of the \textit{Gauss laws}
\begin{equation} \label{eq:Gauss_laws}
    G_a(\bm{x}) \ket{\Psi} = 0 \qquad \forall \bm{x}\in\mathds{Z}^2, a .
\end{equation}
Here, and in the following, we assume the absence of static charges. Since $G_a(\bm{x})$ can be seen as the divergence of electric field values, eq.~\eqref{eq:Gauss_laws} tells us that the ingoing flux at each site (from the left and down links) equals the outgoing one (on the right and upper links). 

\begin{figure}[t]
    \centering
   \includegraphics[width=0.75\textwidth]{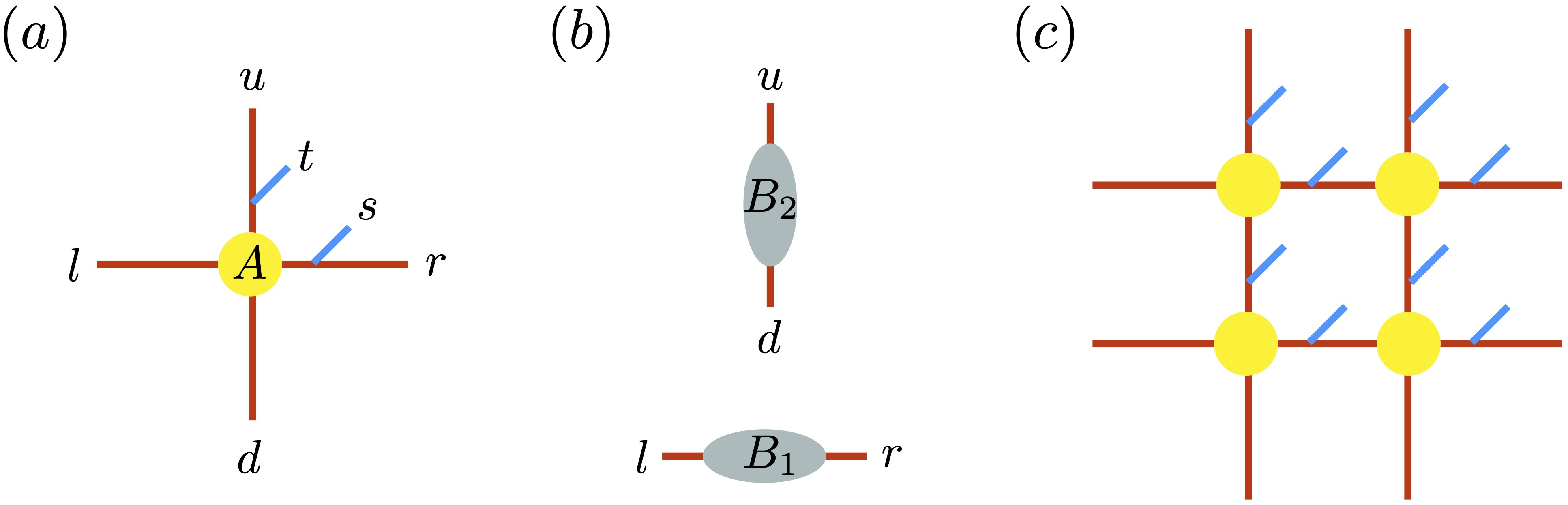}
    \caption{Ingredients of a gauged PEPS tensor network. (a) On each lattice site a tensor $A^{st}_{ruld}$ with 4 virtual indices $r,u,l,d$ and two physical gauge indices $s,t$ is placed. (b) Virtual states on neighboring sites are projected onto maximally entangled states via contraction with tensors $B_{1,2}$ on the links. (c) A PEPS with only physical indices (after projection) is obtained on the lattice.}
    \label{fig:PEPS_tensors}
\end{figure}

~\\
In this paper we make use of the gauged PEPS ansatz of \cite{Zohar:2015jnb}, which allows us to study a pure LGT using a tensor network construction. (For further explorations of this formalism see \cite{Zohar:2015eda,Zohar:2016wcf,Zohar:2017yxl,Emonts:2022yom}.) We assume a 2$D$ lattice with periodic boundary conditions in both directions. On each lattice site a tensor $A^{st}_{ruld}$ is placed. Fig.\,\ref{fig:PEPS_tensors}(a) shows a graphical representation of this tensor using diagrammatic tensor network notation.\,\footnote{For an introduction into this topic we refer to, e.g., \cite{Eisert:2013gpa}.} The tensor $A^{st}_{ruld}$ has 4 \textit{virtual} indices (or \textit{legs}), labelled as $r,u,l,d$ for the respective directions \textit{right}, \textit{up}, \textit{left} and \textit{down}. The $l$ and $d$ legs are considered ingoing, while the $r$ and $u$ legs are the outgoing ones. They are associated with the virtual Hilbert spaces $\mathcal H_r$, $\mathcal H_u$, $\mathcal H_l$ and $\mathcal H_d$, which are spanned by group multiplet states $\ket{jm}$ ($j$ labels irreps of $\mathcal G$, $m$ an identifier within each). The numerical size of these virtual indices is denoted as the bond dimension $\chi$ and allows to vary the number of variational parameters. While all states within a multiplet must be included, the overall number of multiplets may be truncated (for detailed discussions see \cite{Zohar:2015jnb,Zohar:2021wqy}). The additional indices $s$ and $t$ (standing for \textit{side} and \textit{top}) correspond to the \textit{physical} gauge field Hilbert spaces $\mathcal H_s$ and $\mathcal H_t$, placed respectively on the $r$ and $u$ links. As discussed above, they are spanned by the dual representation basis states $\ket{jmn}$, in which the irreps might be truncated as well, that is, a subset of the irreps $j$ of the group is included. The resulting physical-virtual state 
\begin{equation}
    \ket{A(\bm{x})} = A^{j_sm_sn_s;j_tm_tn_t}_{j_rm_r;j_um_u;j_lm_l;j_dm_d} \ket{j_sm_sn_s;j_tm_tn_t} \ket{j_rm_r;j_um_u;j_lm_l;j_dm_d}
\end{equation}
on each lattice site is therefore an element in the Hilbert space $\mathcal H_s \otimes \mathcal H_t \otimes \mathcal H_r \otimes \mathcal H_u \otimes \mathcal H_l \otimes \mathcal H_d$.

The second building block of the PEPS are projector states $B_{1,2}$ on the links (cf.\ Fig.\,\ref{fig:PEPS_tensors}(b)), defined as 
\begin{equation}
    \ket{B_1(\bm{x})} = \sum_j \ket{jm}_{r,\bm{x}} \ket{jm}_{l,\bm{x}+\hat{\bm{e}}_1} ,\quad \ket{B_2(\bm{x})} = \sum_j \ket{jm}_{u,\bm{x}} \ket{jm}_{d,\bm{x}+\hat{\bm{e}}_2} .
\end{equation}
Their purpose is to project the virtual states between two neighboring sites onto maximally entangled states via contraction. In this way, the contraction of all virtual tensor indices over the whole lattice,
\begin{equation}
    \ket{\Psi} = \bigotimes_{\bm{x},i} \bra{B_i(\bm{x})} \bigotimes_{\bm{x}} \ket{A(\bm{x})} ,
\end{equation}
yields a PEPS $\ket{\Psi}$ with only physical (gauge) degrees of freedom, cf.\ Fig.\,\ref{fig:PEPS_tensors}(c).

We assume the tensor network to be translationally invariant, i.e.\ the same tensors are placed on all sites. In order for the resulting PEPS $\ket{\Psi}$ to be gauge invariant, it has to obey the condition \eqref{eq:gauge_inv_state}. Following the construction in \cite{Zohar:2015jnb}, one can directly see that this condition is fulfilled (cf.\ Fig.~\ref{fig:lattice_inv}) if the physical-virtual state satisfies $(\forall \bm{x}\in\mathds{Z}^2, g\in\mathcal{G})$
\begin{align}
    \tilde\Theta_g^s(\bm{x}) \tilde\Theta_g^t(\bm{x}) \ket{A(\bm{x})} &= \theta_g^l(\bm{x}) \theta_g^d(\bm{x}) \ket{A(\bm{x})} ,\nonumber\\
    \Theta_g^s(\bm{x}) \ket{A(\bm{x})} &= \tilde\theta_g^r(\bm{x}) \ket{A(\bm{x})} ,\label{eq:tensor_inv}\\
    \Theta_g^t(\bm{x}) \ket{A(\bm{x})} &= \tilde\theta_g^u(\bm{x}) \ket{A(\bm{x})} ,\nonumber
\end{align}
or graphically
\begin{equation}
    \vcenter{\hbox{\includegraphics[width=\textwidth]{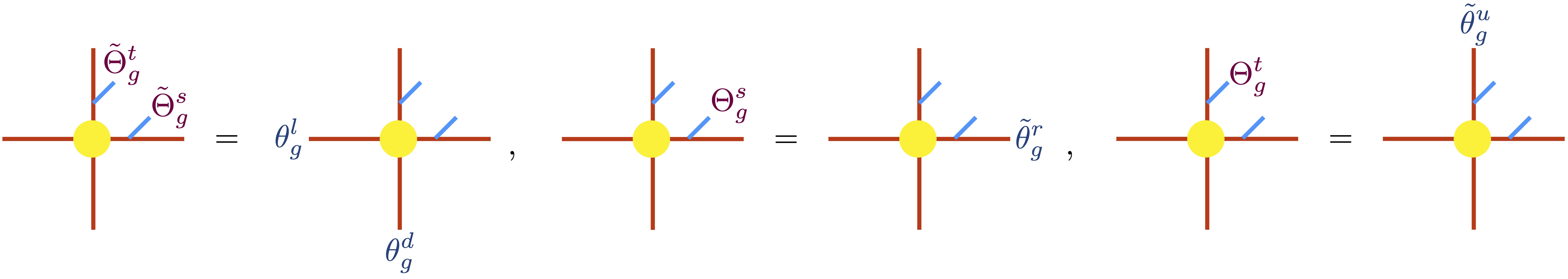}}} ,
\end{equation}
and the projectors obey
\begin{equation} \label{eq:proj_inv}
    \tilde\theta_g^r(\bm{x}) \theta_g^{l\dagger}(\bm{x}+\hat{\bm{e}}_1) \ket{B_1(\bm{x})} = \ket{B_1(\bm{x})} ,\quad \tilde\theta_g^u(\bm{x}) \theta_g^{d\dagger}(\bm{x}+\hat{\bm{e}}_2) \ket{B_2(\bm{x})} = \ket{B_2(\bm{x})} ,
\end{equation}
that is
\begin{equation}
    \vcenter{\hbox{\includegraphics[width=0.5\textwidth]{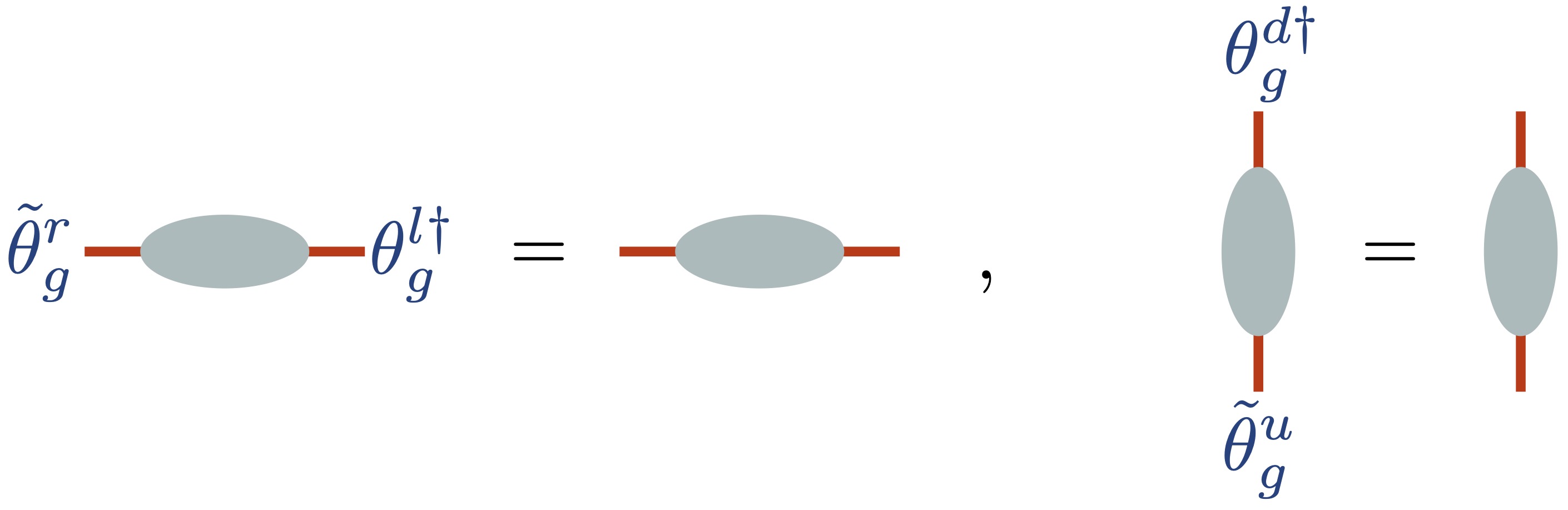}}} .
\end{equation}
In this construction, the physical gauge transformations act as $\Theta_g^p\ket{jmn} = \ket{jmn'}D^j_{nn'}(g)$ and $\tilde\Theta_g^p\ket{jmn} = D^j_{mm'}(g)\ket{jm'n}$ for both $p=s,t$ indices. On the other hand, the virtual Hilbert spaces are transformed as $\theta_g^e\ket{jm} = \ket{jm'}D^j_{m'm}(g)$ and $\tilde\theta_g^e\ket{jm} = D^j_{mm'}(g)\ket{jm'}$ for all $e=r,u,l,d$ indices.

Considering again the case of a compact Lie group, the first condition in \eqref{eq:tensor_inv} implies $j_s \otimes j_t \cong j_l \otimes j_d$, i.e.\ the  congruence between combined physical representations and combined ingoing virtual ones. The remaining two conditions imply $j_s \cong j_r$ and $j_t \cong j_u$, meaning that the individual physical representations are identical to the virtual ones on the same leg.

\subsection{Transfer operators}
\label{subsec:transfer_ops}

The key element for our entanglement calculations in the next section is the so-called transfer operator. The local on-site transfer operator is defined by taking two copies of the physical-virtual tensor and contracting the physical indices between them. This general prescription is valid for any tensor network type. We employ it here to the gauged PEPS framework, for which the local on-site transfer operator can be written as
\begin{equation} \label{eq:T1}
    \hat T^{(1)} = T^{(1)}_{ll',rr',dd',uu'} \ket{ll'}\bra{rr'} \otimes \ket{dd'}\bra{uu'} .
\end{equation}
This tensor $\hat T^{(1)}$ can be interpreted as a map from the ingoing Hilbert spaces (represented by ket vectors on the $l$ and $d$ legs) to the outgoing ones (given as dual bras on the $r$ and $u$ legs). The matrix elements of $\hat T^{(1)}$ are given as
\begin{equation} \label{eq:T1_elem}
    T^{(1)}_{ll',rr',dd',uu'} = \Tr_{s,t}\left[ A^{st}_{ruld}\ket{st}\bra{s't'}\bar A^{s't'}_{r'u'l'd'} \right] = \vcenter{\hbox{\includegraphics[width=0.3\columnwidth]{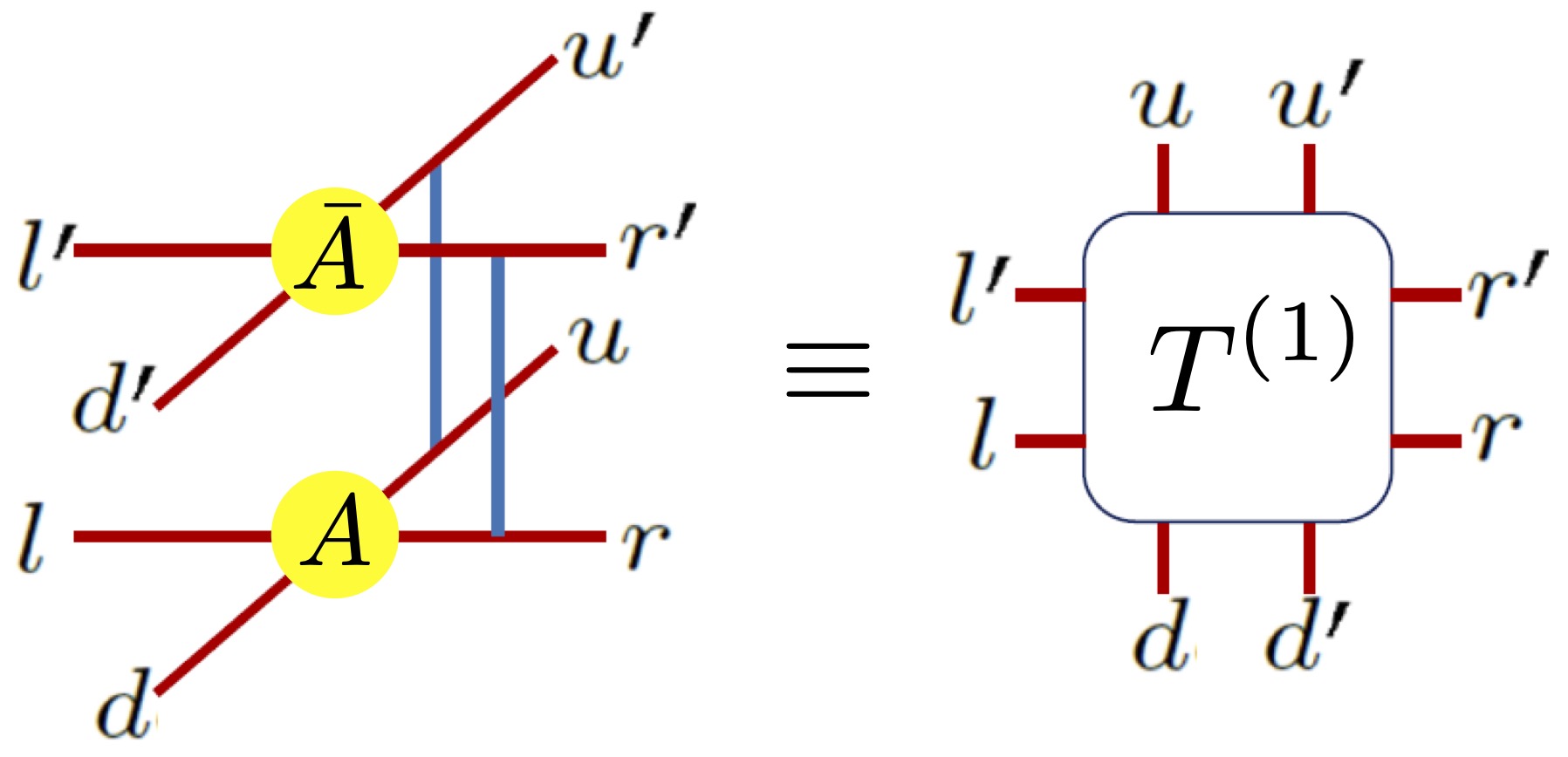}}} .
\end{equation}
Here, two sets of physical indices, $\{s,t\}$ and $\{s',t'\}$, of a physical-virtual PEPS tensor $A^{st}_{ruld}$ and its adjoint $\bar A^{s't'}_{r'u'l'd'}$ are contracted, leaving a tensor with in total 8 indices, $\{ll',rr',dd',uu'\}$.

Let us study properties of the transfer operator when imposing the gauge symmetry. The symmetry constraints \eqref{eq:tensor_inv} of the PEPS tensors imply
\begin{equation} \label{eq:T1_inv}
    \vcenter{\hbox{\includegraphics[width=0.65\textwidth]{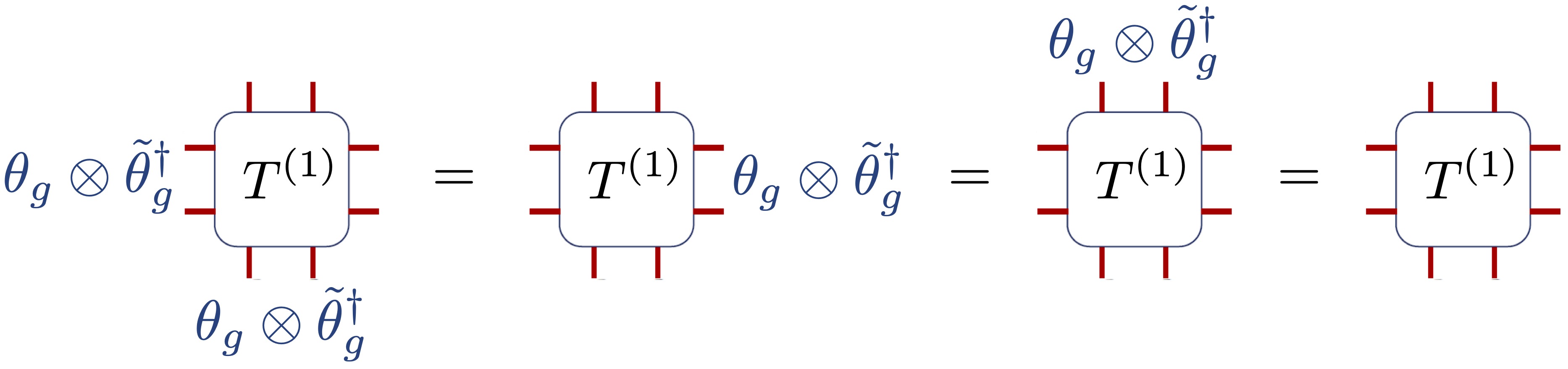}}} .
\end{equation}
That means that vectors on the ingoing legs combine together to a \textit{joint} singlet under $(\theta_g^l \otimes \tilde\theta_g^{l'\dagger}) \otimes (\theta_g^d \otimes \tilde\theta_g^{d'\dagger})$, while vectors on the outgoing ones are separately \textit{on-leg} singlets under the respective transformations $(\theta_g \otimes \tilde\theta_g^{\dagger})$. Symbolically, we can illustrate this map as follows
\begin{equation} \label{eq:T1_map}
    \vcenter{\hbox{\includegraphics[width=0.17\textwidth]{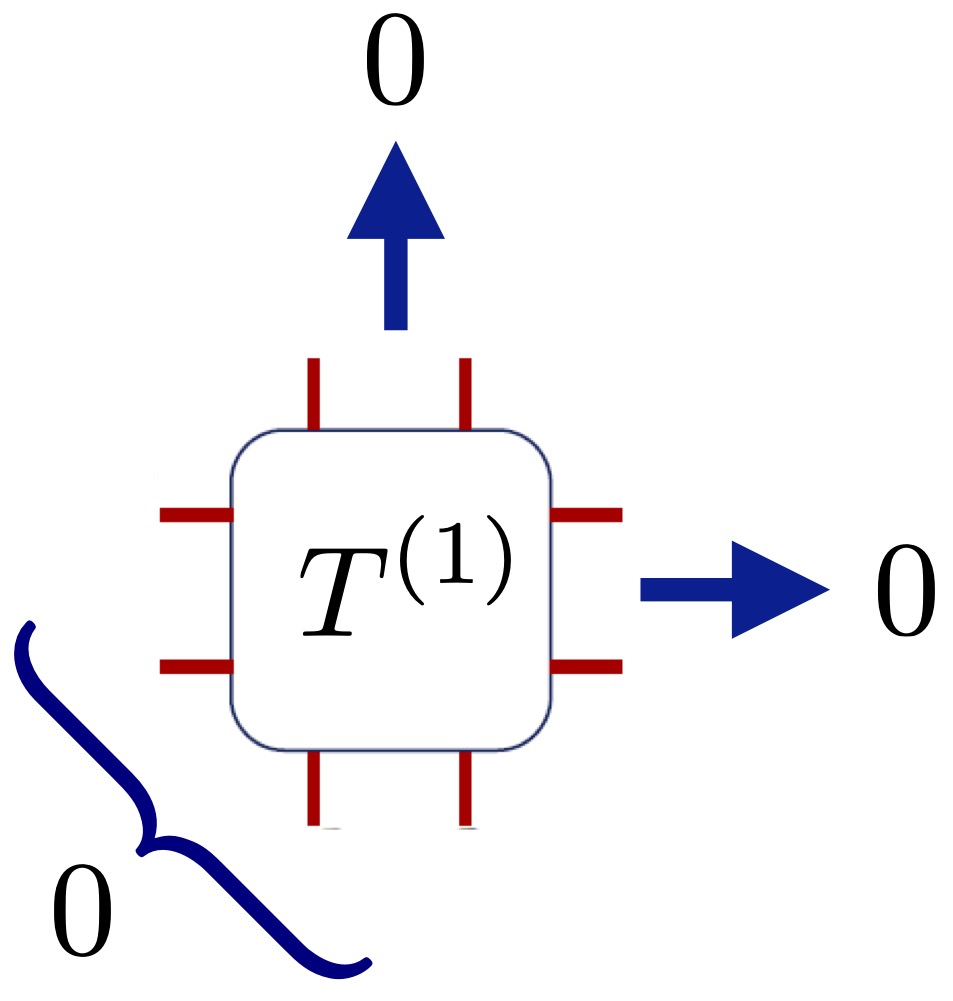}}} .
\end{equation}

When calculating entanglement quantities in the next section, we tile the whole 2$D$ lattice with transfer operator combinations. From \eqref{eq:T1_map} it becomes apparent, that in such a case the input legs ($l,d$) are contracted with on-leg singlets from the right and up index of the neighboring site to the left and below. Defining the projector as $\Pi_0 \equiv \sum_j \ket{0(j)}\bra{0(j)}$, it then suffices to analyze the tensor $\hat\tau_0 = \Pi_0\otimes\Pi_0\hat T^{(1)}$, which takes only on-leg singlets $\ket{0(j)} \equiv \ket{jmjm}$ as inputs. This version of the transfer operator will be the central building block in the next section. It can be written as
\begin{equation}
    \hat\tau_0 = \sum_{\{j\}} (\tau_0)_{j_l,j_r;j_d,j_u} \ket{0(j_l)}\bra{0(j_r)} \otimes \ket{0(j_d)}\bra{0(j_u)} .
\end{equation}
As shown in \cite{Zohar:2021wqy}, $(\tau_0)_{j_l,j_r;j_d,j_u}$ is a positive symmetric matrix, for which a decomposition
\begin{equation} \label{eq:tau0_VLV}
    \tau_0 = V \Lambda V^\dagger
\end{equation}
with orthogonal $V$ and diagonal $\Lambda$ (containing the eigenvalues $\lambda_\mu$) is existing. The operator $\hat\tau_0$ hence can be written as 
\begin{equation} \label{eq:tau0_Mform}
    \hat\tau_0 = \sum_\mu \lambda_\mu \hat M_\mu \otimes \hat M_\mu ,
\end{equation}
where the $\hat M_\mu$ are given as $\hat M_\mu = \sum_{j_1,j_2} V_{j_1 j_2 \mu} \ket{0(j_1)}\bra{0(j_2)}$. In \eqref{eq:tau0_Mform}, these act independently on the horizontal ($l,r$) and vertical indices ($d,u$).

\section{Area laws for R\'enyi entanglement entropies}
\label{sec:area}

\subsection{Idea in 1$D$}
\label{sec:idea}

In this section, we introduce the underlying idea of a transfer operator approach, which allows us to calculate normalized R\'enyi entanglement entropies of order $n>1$, defined as
\begin{equation} \label{eq:S_n_norm}
    \bar S_n = \frac{1}{1-n} \ln\left( \frac{\Tr[\rho_A^n]}{\Tr[\rho_A]^n} \right) .
\end{equation}
Here, the reduced density matrix is defined as $\rho_A = \Tr_B[\ket{\Psi}\bra{\Psi}]$ for a pure quantum state $\ket{\Psi}$. Contrary to the usually defined R\'enyi entanglement entropy, 
\begin{equation} \label{eq:S_n_reg}
    S_n = \frac{1}{1-n} \ln\left( \Tr[\rho_A^n] \right) , 
\end{equation}
we assume an additional normalization in \eqref{eq:S_n_norm}, allowing us to simplify the construction for the subsequent PEPS framework. Note that in the limit $n \rightarrow 1^+$, $S_n$ approaches the entanglement entropy,
\begin{equation} \label{eq:S_1}
    S_1 = -\Tr[\rho_A \ln \rho_A] ,
\end{equation}
defined as the von Neumann entropy of the reduced density matrix.

The only nontrivial element in the definition \eqref{eq:S_n_norm} is the trace in the numerator. We demonstrate this calculation for the case $n=2$. For a 1$D$ QMB state, which is expressed in MPS form (and without assuming any gauge invariance), the so-called purity
\begin{equation} \label{eq:p_2}
    p_2 \equiv \Tr[\rho_A^2]
\end{equation}
takes the form
\begin{equation} \label{eq:p2_MPS}
    p_2 \quad=\quad \vcenter{\hbox{\includegraphics[width=0.6\textwidth]{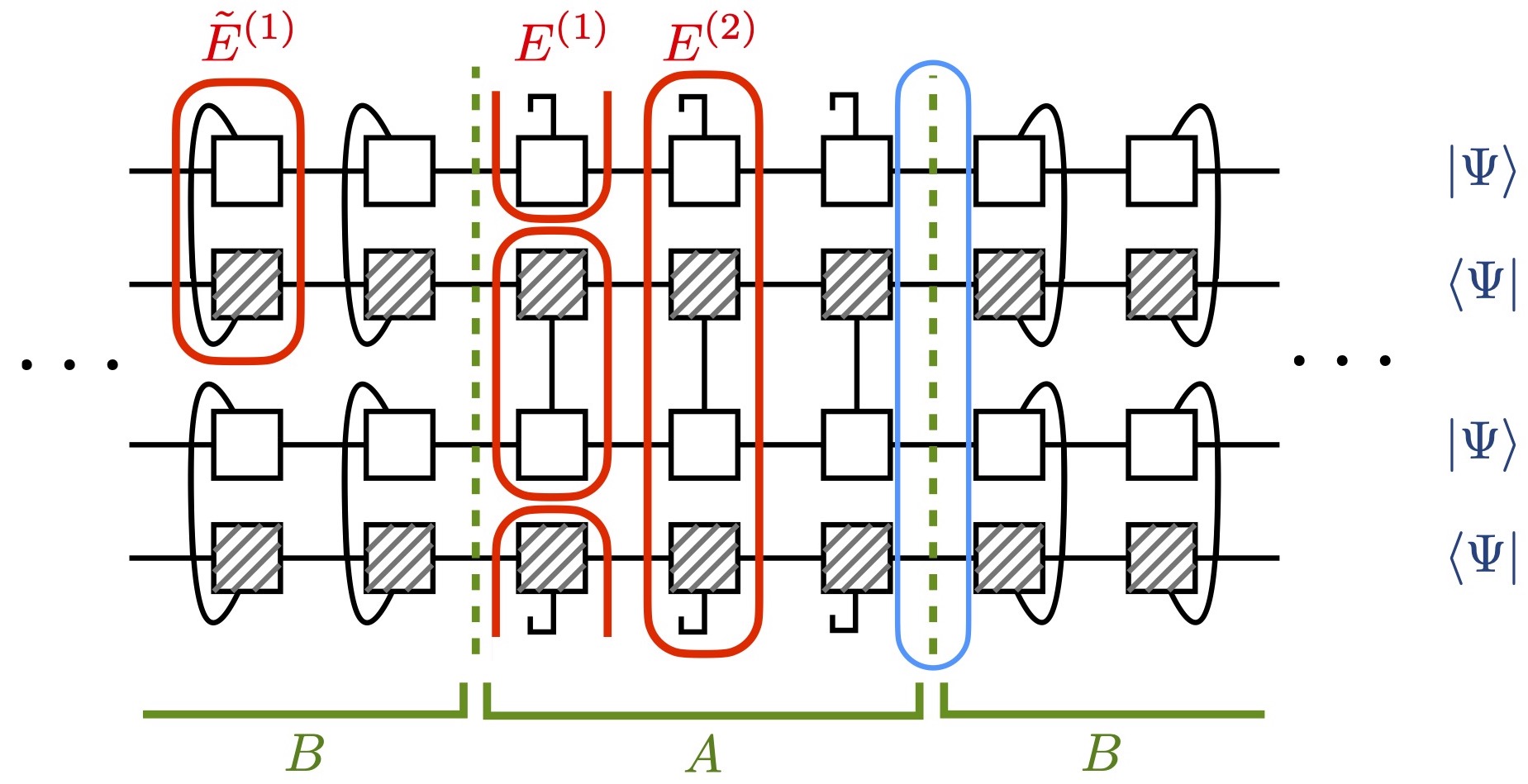}}} .
\end{equation}
In eq.\,\eqref{eq:p2_MPS}, we make use of diagrammatic MPS notation.\,\footnote{Specifically for a MPS $\ket{\Psi}$, rank-3 tensors $A^{(k)}_{\alpha\beta i_k}$ are placed at each lattice site $k=1,\ldots,N$. Such a tensor is graphically represented by a box with 3 legs. The horizontal legs correspond to the \textit{virtual indices} $\alpha,\beta$, whose size is denoted as the \textit{bond dimension} $\chi$. The leg pointing upwards denotes the \textit{physical index} $i_k$. The pairwise contraction of all virtual indices along the chain yields the physical quantum state.}
Each row corresponds to the alternating ket- and bra-tensors of the quantum state. We assume an infinitely large, periodic chain and exemplarily defined the subsystem $A$ as the 3 tensors in the middle, such that the complement $B$ is given as the outside region (cf.\ green marks). Different types of traces in both regions are indicated by connected physical indices of MPS tensors. The crucial observation in this expression is the following. The quantity of interest is uniformly composed out transfer operators. Inside the subsystem $A$ in the tensor network diagram in \eqref{eq:p2_MPS}, we have marked a single-site transfer operator $E^{(1)}$ by a small red box. In fact, this tensor appears twice, such that one can write an entire column as $E^{(2)} \equiv E^{(1)} \otimes E^{(1)}$ (indicated by a large red box). In the complement region $B$, we have marked a repeating tensor $\tilde E^{(1)}$, which is marginally different from $E^{(1)}$ by the ordering of indices. 

A nontrivial tensor contraction appears only at the two subsystem boundaries, marked by green dashed lines in \eqref{eq:p2_MPS}. Let us consider exemplarily the right one between subsystem $A$ and the complement $B$ (cf.\ the blue box). This part of the tensor network diagram can be simplified as follows:
\begin{equation} \label{eq:p2_bdy}
    \vcenter{\hbox{\includegraphics[width=0.95\textwidth]{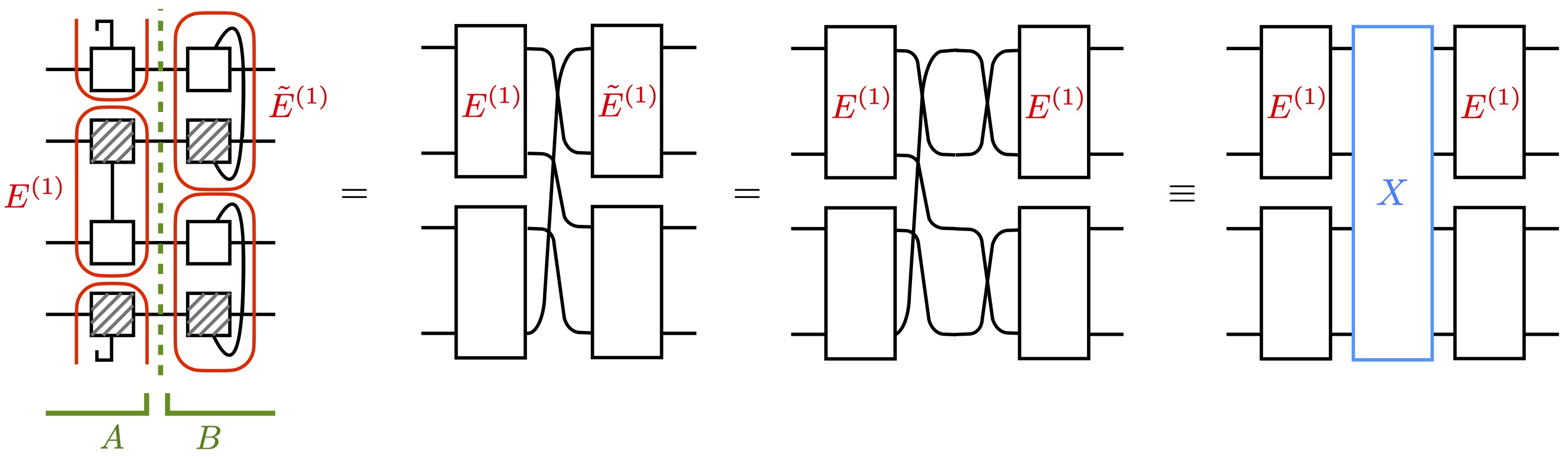}}} .
\end{equation}
For the first equality in \eqref{eq:p2_bdy} we have rewritten the expression in terms of $E^{(1)}$ and $\tilde E^{(1)}$. In the next equality we made use of the fact that $\tilde E^{(1)}$ is given as $\tilde E^{(1)} = P E^{(1)} P$, where $P$ is the operator permuting the two tensor indices on each side of the transfer operator. We finally defined the boundary operator $X$, which realizes the shifting and permutation of tensor indices in-between two copies of $E^{(2)}$. If we label the tensor indices from top to bottom as $1,\ldots,4$, $X$ can be effectively seen as the following permutation operator
\begin{equation}
    X: \{ 1,2,3,4 \} \mapsto \{ 1,4,3,2 \} .
\end{equation}
One easily finds that the left boundary of the subsystem takes the same form. Since all remaining permutation operators to the left and right side cancel each other, the purity can be re-expressed entirely in terms of $E^{(2)}$ and two insertions of boundary operators $X$,
\begin{equation}
    p_2 \quad=\quad \vcenter{\hbox{\includegraphics[width=0.7\textwidth]{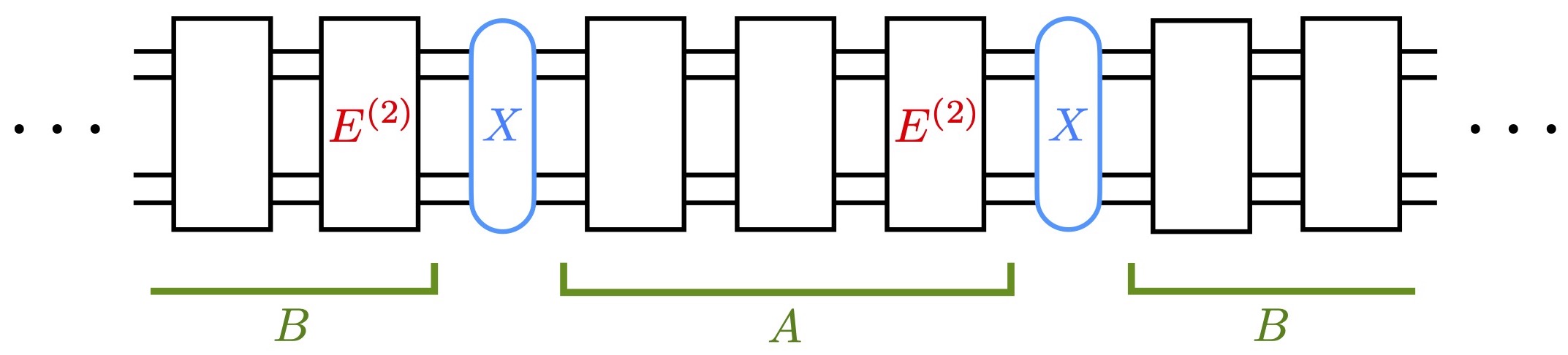}}} . 
\end{equation}
This expression, and its generalization to higher orders $n$, allows us to calculate the desired R\'enyi entanglement entropies in \eqref{eq:S_n_norm}.

\subsection{Setup in 2$D$}

\subsubsection{Contraction of transfer operator plaquettes}

The previously introduced transfer operator method can be directly generalized to two dimensions using PEPS. We explain this method again for order $n=2$ here. As it became apparent from the MPS construction, the essential tensors for calculating the purity are a double copy of the transfer operator and the boundary operator. The structure of the latter is unaltered. For the former, we make use of the operator $\hat\tau_0$, which takes the gauge symmetries into account (cf.\ section~\ref{subsec:transfer_ops}), and define a single-site double copy of the transfer operator as
\begin{equation} \label{eq:T2}
    \mathcal T^{(2)} \ \vcenter{\hbox{\includegraphics[width=0.35\columnwidth]{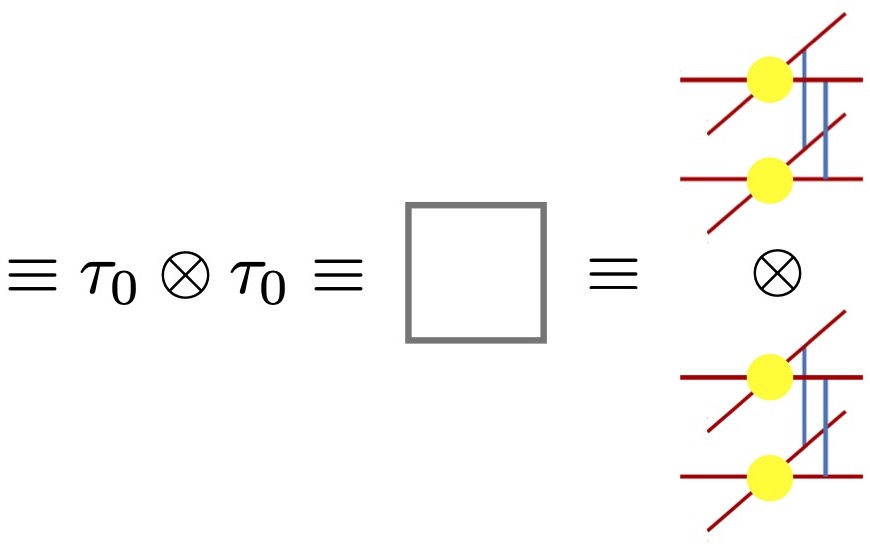}}} .
\end{equation}
Here, we have introduced a new plaquette notation for $\mathcal T^{(2)}$, represented by a gray square. 

\begin{figure}[t]
    \centering
   \includegraphics[width=0.49\textwidth]{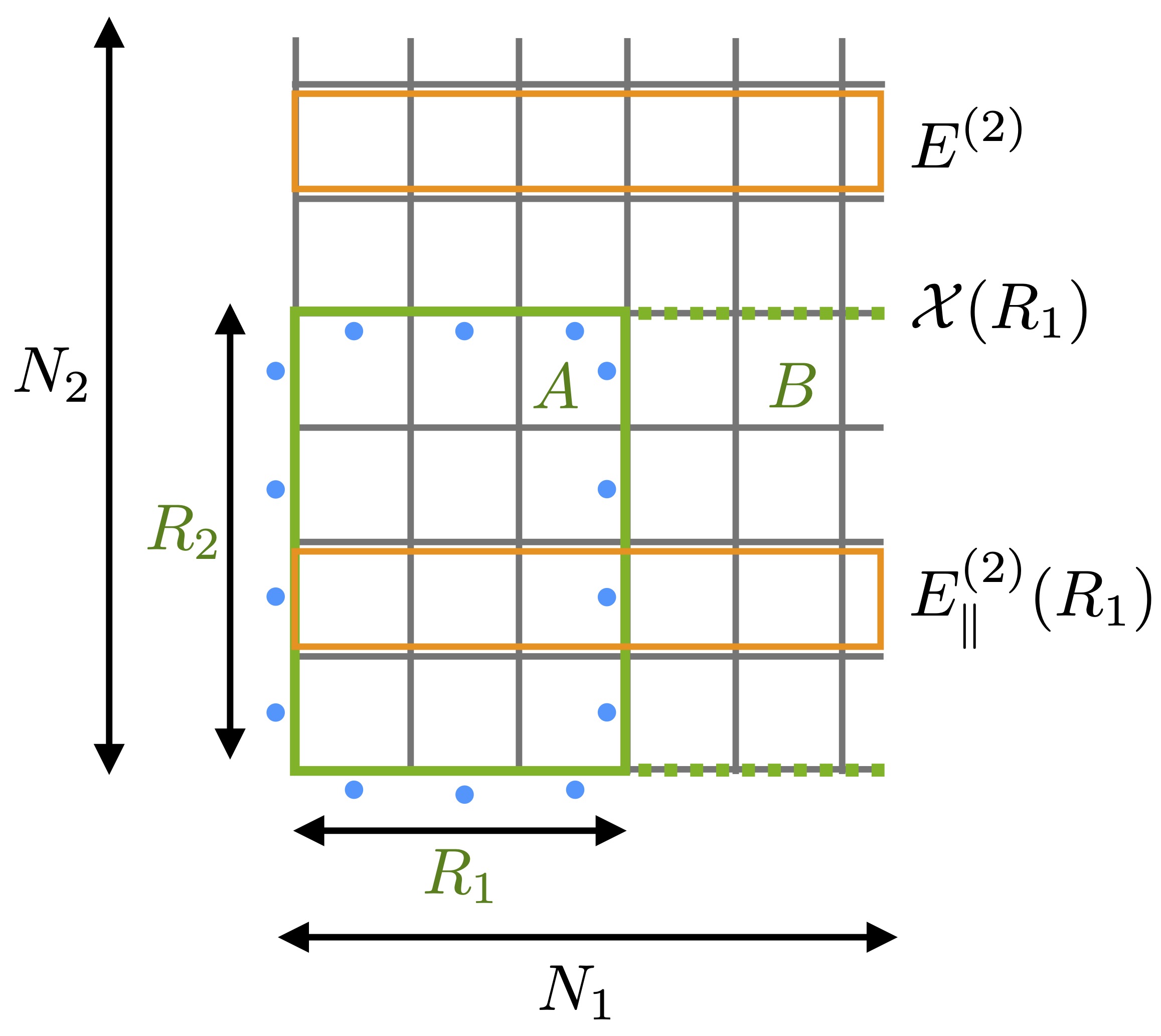}
    \caption{Setup for the calculation of the purity and R\'enyi entanglement entropy for a LGT on a 2$D$ lattice. The lattice is tiled with transfer operator plaquettes (gray squares) defined in \eqref{eq:T2}. A rectangular subregion $A$ is chosen in the lower left corner (green frame). The purity \eqref{eq:p2_PEPS} is calculated by contracting different types of transfer rows (orange boxes) and boundary operators (green dashed lines). Blue dots indicate the position of gauge fields at the boundary of $A$. See text for further explanations.}
    \label{fig:setup}
\end{figure}

When calculating the purity and subsequently other entanglement measures for a LGT on a 2$D$ lattice, we have to define $\mathcal T^{(2)}$ for each lattice site. The resulting tensor network setup is illustrated in Fig.~\ref{fig:setup}. We assume a periodic system of $N_1 \times N_2$ lattice sites $\bm{x} \equiv (x,y)$ in the horizontal and vertical direction, respectively. The lattice is tiled with transfer operator plaquettes $\mathcal T^{(2)}$. In general, we assume that a subsystem $A$ can be defined as any connected region within this 2$D$ lattice, which includes only full plaquettes. In Fig.~\ref{fig:setup}, we have placed a rectangular subregion $A$ in the lower left corner of the lattice (marked as a green frame). The subsystem has size $R_1 \times R_2$, where we measure the dimensions $R_{1,2}$ in units of the lattice spacing $a$.\,\footnote{Note that we define $N_{1,2}$ to be an integer number, while $R_{1,2}$ is dimensionful. The subregion $A$ hence includes $R_{1,2}/a$ plaquettes in each direction. When deriving the final result in the thermodynamic limit, this convention will turn out to be convenient.} Along the boundary of $A$, the boundary tensor $X$ is inserted at each lattice site. 

To calculate $p_2$ one needs to contract all tensors in both dimensions. This procedure can be done either column- or row-wise. For the following evaluation we choose to calculate transfer rows by contracting tensors in the horizontal direction first. The transfer rows are afterwards contracted in the vertical direction from bottom to top. This yields the following expression for the purity
\begin{equation} \label{eq:p2_PEPS}
    p_2 = \Tr\left[\mathcal X(R_1) E^{(2)R_2/a}_{||}(R_1) \mathcal X(R_1) E^{(2) N_2-R_2/a}\right] .
\end{equation}
In eq.~\eqref{eq:p2_PEPS}, we have defined three types of transfer rows, which are visualized in Fig.~\ref{fig:setup}. In the bottom row, the boundary operator $X$ is inserted at each lattice site $0 < x < R_1$ along the boundary of $A$. We define this operator as
\begin{equation} \label{eq:X}
\mathcal X(R_1) \equiv X(x=a) \otimes \ldots \otimes X(x=R_1) \otimes \mathds{1}(x=R_1+a) \otimes \ldots \otimes \mathds{1}(x=N_1 a) ,
\end{equation}
where outside the subsystem ($x>R_1$) only identity operators are inserted (indicated by a green dashed line in Fig.~\ref{fig:setup}). The rows above, which are inside the subsystem $A$, include $\mathcal T^{(2)}$ at each site and one insertion of $X$ at the left and right boundary. We denote the corresponding transfer operator as $E^{(2)}_{||}$, taking the form
\begin{equation} \label{eq:E2p}
    E_{||}^{(2)}(R_1) = \Tr_{\text{row}}\left[X \otimes \mathcal T^{(2)\otimes R_1/a} \otimes X \otimes \mathcal T^{(2) \otimes N_1-R_1/a}\right] . 
\end{equation}
$E_{||}^{(2)}$ depends on $R_1$ through the second insertion of $X$ at position $x=R_1$. 
After one additional insertion of $\mathcal X(R_1)$ at the top of the subregion, the tiling of the 2$D$ lattice is completed by the remaining copies of the transfer row $E^{(2)}$ consisting solely out of single-site transfer operators,
\begin{equation} \label{eq:E2}
    E^{(2)} = \Tr_{\text{row}}\left[\mathcal T^{(2)\otimes N_1} \right] .
\end{equation}
Formula~\eqref{eq:p2_PEPS} is derived by taking $R_2/a$ copies of $E_{||}^{(2)}$ and $N_2-R_2/a$ copies of $E^{(2)}$ to tile the whole lattice.

The normalization $\Tr[\rho_A]^2$ in \eqref{eq:S_n_norm} can be easily included by observing that $\Tr[\rho_A]^2 \equiv \Tr_A[\Tr_B[\rho]]^2 = \Tr[\rho]^2$. Defining a single-copy transfer row identically to the Wilson loop studies in \cite{Zohar:2021wqy} as
\begin{equation} \label{eq:E1}
    E^{(1)} = \Tr_{\text{row}}[\tau_0^{\otimes N_1}] ,
\end{equation}
we have
\begin{equation}
    \Tr[\rho_A]^2 = \Tr\left[ E^{(1) N_2} \right]^2 .
\end{equation}

\subsubsection{Hilbert space decomposition and Gauss laws}

As alluded in the introduction, our transfer operator approach implicitly assumes an extended Hilbert space construction \cite{Buividovich:2008gq,Donnelly:2011hn,Ghosh:2015iwa,Aoki:2015bsa,Soni:2015yga} for the calculation of the normalized R\'enyi entanglement entropies. In this framework, the tensor product of the gauge field Hilbert spaces $\mathcal H_\mathtt{l}$ on all links $\mathtt{l}$ forms an embedding for gauge-invariant states. A partial trace over the link Hilbert spaces belonging to the complement region $B$ is well-defined, such that the reduced density matrix can be calculated and the system admits the desired decomposition
\begin{equation}
    \mathcal H_{ext} \equiv \bigotimes_\mathtt{l} \mathcal H_\mathtt{l} = \mathcal H_A \otimes \mathcal H_B .
\end{equation}

Apart from the Hilbert space decomposition, it is important to analyze the gauge invariance for this approach, which is encapsulated in the Gauss laws. In the context of entanglement measures, the relevant quantity is the reduced density matrix $\rho_A$. We want to exemplarily discuss gauge invariance of $\rho_A$ for the Abelian $U(1)$ group in this section. Naturally, modifications of the Gauss laws can be expected only at the subsystem boundary. Moreover, lattice points in $B$, which are not neighboring with $A$, are not relevant for the following discussion. 

\begin{figure}[t]
    \centering
   \includegraphics[width=0.8\textwidth]{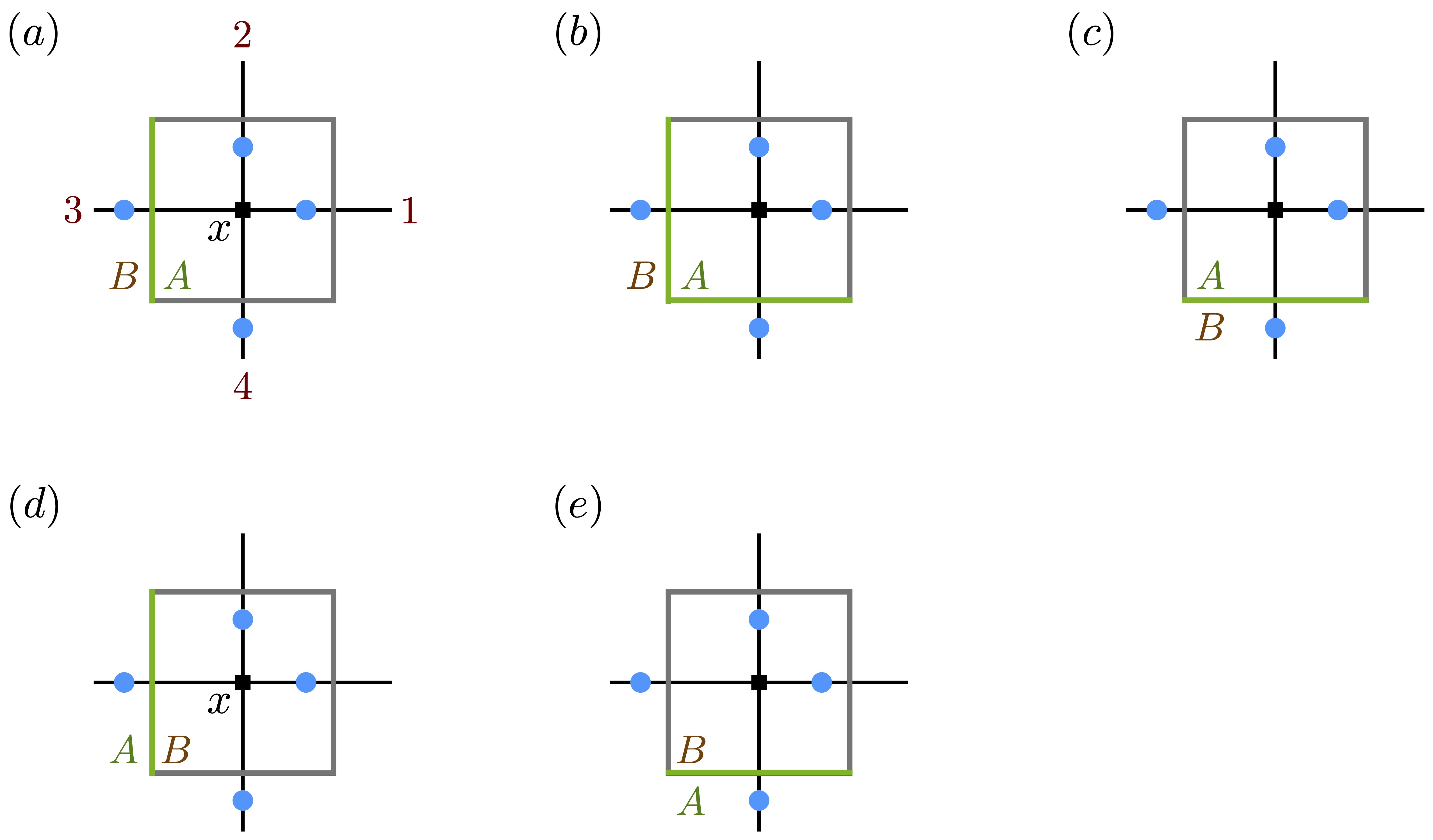}
    \caption{Modifications of the Gauss laws along the subsystem boundary: The transfer operator approach is based on an extended Hilbert space formalism, which allows to realize a Hilbert space decomposition. Panels (a-e) illustrate all different possibilities of lattice sites along the boundary of the subsystem $A$. In these cases, the reduced density matrix $\rho_A$ is gauge transformed according to \eqref{eq:Gausslaw_mod} for an position-dependent effective Gauss law operator $\widetilde G(\bm{x})$. Blue dots indicate gauge fields on links around a lattice site $\bm{x}$. Green lines mark the subsystem boundary to the complement region $B$. A gray frame indicates a single transfer operator plaquette (cf.\ also Fig.\,\ref{fig:setup}). See text for further discussions.}
    \label{fig:Gausslaws}
\end{figure}

The precise setting of the link decomposition follows from the plaquette tiling of transfer operators. As defined in the previous section and illustrated in Fig.~\ref{fig:setup}, the subsystem $A$ consists of full plaquettes. A single transfer operator $\tau_0$ and its double copy $\mathcal T^{(2)}$ in \eqref{eq:T2} contain the gauge fields on the right and up leg. As a consequence, all four gauge fields around any lattice site $\bm{x} \in A \subset \mathds{Z}^2$, which is not at the left or bottom boundary of the subsystem, are completely inside $A$. For those lattice points, the reduced density matrix is invariant under
\begin{equation}
    \e^{i\phi G(\bm{x})} \rho_A \e^{-i\phi G(\bm{x})} = \rho_A ,
\end{equation}
where the Gauss law operator (in absence of charges) takes the standard form
\begin{equation} \label{eq:G_x}
    G(\bm{x}) = E_1 + E_2 - E_3 - E_4 .
\end{equation}
Here, we have labelled the electric field variables $E_i$ as indicated in Fig.~\ref{fig:Gausslaws}(a).

The first exception to this rule appears for lattice sites inside $A$ at the left and bottom boundary of the subsystem. This is due to the fact that at these locations, the gauge fields around a lattice site belong to different regions. (Compare Fig.~\ref{fig:setup}, where we have marked the gauge fields along the boundary of $A$ by blue dots.) At these parts of the boundary, three different types of exceptions arise, which are illustrated in Fig.~\ref{fig:Gausslaws}(a-c). In case (a) the subsystem boundary is at the left side of the plaquette, hence the third gauge field belongs to region $B$, while all others are inside $A$. Similarly, in case (b) for the lower left corner, $E_3$ and $E_4$ are outside of $A$, while for case (c) only $E_4$ is inside $B$. The invariance properties of $\rho_A$ at these points can be derived by using the gauge invariance of the full density operator $\rho = \ket{\Psi}\bra{\Psi}$, given as $\e^{i\phi G(\bm{x})}\rho \e^{-i\phi G(\bm{x})} = \rho$ w.r.t.\ the full Gauss law operator \eqref{eq:G_x} ($\forall \bm{x} \in \mathds{Z}^2$). In case (a) we therefore have
\begin{align}
    \rho_A &\equiv \Tr_B[\rho] = \Tr_B[\e^{i\phi G(\bm{x})}\rho \e^{-i\phi G(\bm{x})}] \nonumber\\
    &= \e^{i\phi(E_1+E_2-E_4)} \Tr_B[ \e^{i\phi(-E_3)}\rho \e^{-i\phi(-E_3)} ] \e^{-i\phi(E_1+E_2-E_4)} \nonumber\\
    &= \e^{i\phi(E_1+E_2-E_4)} \rho_A \e^{-i\phi(E_1+E_2-E_4)} \nonumber\\
    &\equiv \e^{i\phi\widetilde G(\bm{x})} \rho_A \e^{-i\phi\widetilde G(\bm{x})} \label{eq:Gausslaw_mod}.
\end{align}
In the second line we have used the fact that only $E_3$ is inside $B$, such that the other electric field values can be pulled out of the trace. In the following line we made use of the cyclicity of the trace, canceling the phase factors. The final expression in eq.~\eqref{eq:Gausslaw_mod} takes the usual form of a gauge invariant operator, but w.r.t.\ the newly defined \textit{effective Gauss law operator} $\widetilde G(\bm{x}) = E_1 + E_2 - E_4$. 
Using the same methodology, we immediately deduce that for case (b) the gauge transformation takes the same form with an effective Gauss law operator given as $\widetilde G(\bm{x}) = E_1+E_2$, while in case (c) we have $\widetilde G(\bm{x}) = E_1+E_2-E_3$.

The second type of Gauss law modifications appears for lattice sites along the top and right subsystem boundary. Here, only the neighboring points in the complement region $\bm{x} \in B$ are affected. Panel (d) and (e) in Fig.~\ref{fig:Gausslaws} show the two possibilities. The gauge transformation of $\rho_A$ is again given by the structural form \eqref{eq:Gausslaw_mod}. In case (d) (representing the right boundary), $\widetilde G(\bm{x})$ follows as $\widetilde G(\bm{x}) = -E_3$, and in case (e) (representing the top boundary) one has $\widetilde G(\bm{x}) = -E_4$.  

In summary, we have discussed the principles of gauge invariance in the proposed transfer operator approach, which is based on an extended Hilbert space formalism. We have shown how the gauge invariance of the full density matrix, given by the regular Gauss laws (for all lattice sites), implies modified Gauss laws for the reduced density matrix at lattice sites along the subsystem boundary. The latter can be represented using an effective Gauss law operator. For notational convenience we have chosen the $U(1)$ gauge group in the previous discussion. The framework, however, can be directly generalized to arbitrary other gauge groups.

\subsection{Results in the thermodynamic limit and continuum}
\label{sec:thermolim}

In this section we derive long-range properties of the normalized R\'enyi entanglement entropies \eqref{eq:S_n_norm} using our transfer operator setup. The results are obtained by considering the thermodynamic limit 
\begin{equation} \label{eq:TD_limit}
    1 \ll R_1/a,R_2/a \ll N_1,N_2 
\end{equation}
for a large subsystem within an infinite lattice. Analogous to the Wilson loop analyses in \cite{Zohar:2021wqy}, we make use of an eigenvalue decomposition of the individual transfer matrices, defined as
\begin{align}
    E^{(1)} &= \sum_i \rho^{(1)}_i \ket{v^{(1)}_i} \bra{w^{(1)}_i} \\
    E^{(2)} &= \sum_i \rho^{(2)}_i \ket{v^{(2)}_i} \bra{w^{(2)}_i} \\
    E^{(2)}_{||}(R) &= \sum_i \rho'^{(2)}_i(R) \ket{v'^{(2)}_i(R)} \bra{w'^{(2)}_i(R)} .
\end{align}
In these expressions we have considered the diagonalization of all transfer matrices $E^{(1)}$, $E^{(2)}$, $E^{(2)}_{||}$, and denoted their eigenvalues respectively as $\rho^{(1)}_i, \rho^{(2)}_i, \rho'^{(2)}_i$. The corresponding right eigenvectors are given by the kets $\ket{v^{(1)}_i}, \ket{v^{(2)}_i}, \ket{v'^{(2)}_i}$, while the associated bras (labelled as $w$) denote the corresponding left eigenvectors. Since $E^{(2)}_{||}(R)$ spatially depends on the insertion of the boundary operator, its eigenvalues and -vectors also depend on $R$.
By construction, the eigenvalues and -vectors of $E^{(2)}$ are related to those of $E^{(1)}$ via
\begin{equation} \label{eq:rho1rho2_rel}
    \rho^{(2)}_i = (\rho^{(1)}_i)^2 ,\qquad\text{and}\qquad \ket{v^{(2)}_i} = \ket{v^{(1)}_i} \otimes \ket{v^{(1)}_i} .
\end{equation}
We assume all eigenvalues to be ordered decreasingly, and the existence of a spectral gap with $K$ ($K'$) degenerate largest eigenvalues of $E^{(2)}$ ($E^{(2)}_{||}$), i.e.\ $|\rho_1^{(2)}| =\ldots= |\rho_K^{(2)}| > |\rho_{K+1}^{(2)}| \ge \ldots$, and $|\rho'^{(2)}_1(R)| =\ldots= |\rho'^{(2)}_{K'}(R)| > |\rho'^{(2)}_{K'+1}(R)| \ge \ldots$.

Using these diagonal forms of the transfer matrices, we can rewrite the purity \eqref{eq:p2_PEPS} as
\footnotesize
\begin{align}
    p_2 = \dim^{4(R_1+R_2)/a}(J)\ \Tr\Biggl[ &\mathcal X(R_1) \Biggl( \rho'^{(2)R_2/a}_1(R_1) \sum_{j=1}^{K'} \ket{v'^{(2)}_j(R_1)}\bra{w'^{(2)}_j(R_1)} + \nonumber\\
    &\phantom{xxxxxx} \sum_{j>K'}\left(\frac{\rho'^{(2)}_j(R_1)}{\rho'^{(2)}_1(R_1)}\right)^{R_2/a} \ket{v'^{(2)}_j(R_1)}\bra{w'^{(2)}_j(R_1)} \Biggr) \nonumber\\
    &\mathcal X(R_1) \Biggl( \rho^{(2)N_2-R_2/a}_1 \sum_{i=1}^{K} \ket{v^{(2)}_i}\bra{w^{(2)}_i} + \sum_{i>K}\left(\frac{\rho^{(2)}_i}{\rho^{(2)}_1}\right)^{N_2-R_2/a} \ket{v^{(2)}_i}\bra{w^{(2)}_i} \Biggr) \Biggr] . \label{eq:p2_PEPS_diag}
\end{align}
\normalsize
In this expression, we have factored out the (possibly degenerate) dominant eigenvalues $\rho'^{(2)}_1$ and $\rho^{(2)}_1$ of the two relevant types of transfer rows. For large loops, as assumed in the thermodynamic limit \eqref{eq:TD_limit}, the second sums containing the subdominant eigenvalues in \eqref{eq:p2_PEPS_diag} vanish. 
Moreover, eq.~\eqref{eq:p2_PEPS_diag} includes a factor $\dim^{4(R_1+R_2)/a}(J)$, which originates from the number of contracted indices along the subsystem boundary, each taking $\dim(J)$ values. As argued in \cite{Zohar:2021wqy}, the singular values depend only on the irrep $J$, but not on these indices. Hence all $\dim(J)$ are equal and contribute only the overall factor to the result.

Let us now include the normalization factor by defining the normalized purity
\begin{equation}
    \bar p_2 \equiv \frac{\Tr[\rho_A^2]}{\Tr^2[\rho_A]} .
\end{equation}
Using \eqref{eq:rho1rho2_rel}, we observe that $(\Tr_A\rho_A)^2 = (\Tr\rho)^2 \overset{N_2\to\infty}{\longrightarrow} (K \rho^{(1)N_2}_1)^2 = K^2 \rho^{(2)N_2}_1$, which cancels the dependence on $N_2$ in \eqref{eq:p2_PEPS_diag}. Employing all simplifications and taking the overall trace, we are left with
\footnotesize
\begin{equation} \label{eq:p2bar_PEPS_row}
    \bar p_2 = \frac{\dim^{4(R_1+R_2)/a}(J)}{K^2} \left(\frac{\rho'^{(2)}_1(R_1)}{\rho^{(2)}_1}\right)^{R_2/a} \sum_{i=1}^{K} \sum_{j=1}^{K'} \bra{w^{(2)}_i}\mathcal{X}(R_1)\ket{v'^{(2)}_j(R_1)} \bra{w'^{(2)}_j(R_1)}\mathcal{X}(R_1)\ket{v^{(2)}_i} .
\end{equation}
\normalsize

To further simplify the resulting equation for $\bar p_2$, we make the important observation that we alternatively, and fully equivalently, could have tiled our lattice column-wise, instead of the row-wise contraction scheme. Under this assumption, $\bar p_2$ is calculated as (compare with eq.~\eqref{eq:p2_PEPS})
\begin{equation}
    \bar p_2 = \frac{\Tr\left[\mathcal X(R_2) E^{(2)R_1/a}_{||}(R_2) \mathcal X(R_2) E^{(2) N_1-R_1/a}\right]}{\Tr^2[E^{(1)N_1}]} .
\end{equation}
Here, $\mathcal{X}(R_2)$, $E^{(2)}_{||}(R_2)$, $E^{(2)}$ and $E^{(1)}$ are calculated as defined in eqs.~\eqref{eq:X}, \eqref{eq:E2p}, \eqref{eq:E2} and \eqref{eq:E1}, respectively, but under the replacements $N_1 \mapsto N_2, R_1 \mapsto R_2$ and a column-wise contraction (i.e.\ trace). Following the same steps that lead to \eqref{eq:p2bar_PEPS_row}, we get
\footnotesize
\begin{equation} \label{eq:p2bar_PEPS_column}
    \bar p_2 = \frac{\dim^{4(R_1+R_2)/a}(J)}{K^2} \left(\frac{\rho'^{(2)}_1(R_2)}{\rho^{(2)}_1}\right)^{R_1/a} \sum_{i=1}^{K} \sum_{j=1}^{K'} \bra{w^{(2)}_i}\mathcal{X}(R_2)\ket{v'^{(2)}_j(R_2)} \bra{w'^{(2)}_j(R_2)}\mathcal{X}(R_2)\ket{v^{(2)}_i} . 
\end{equation}
\normalsize
The result (given as the overall trace for the purity) is invariant under the order of the contraction scheme, hence formulas \eqref{eq:p2bar_PEPS_row} and \eqref{eq:p2bar_PEPS_column} have to be identical. In Appendix~\ref{app:proof} we prove that this condition can be only fulfilled if the dominant eigenvalue $\rho'^{(2)}_1(R)$ has, in the most general case, the spatial dependence
\begin{equation} \label{eq:rho1_form}
    \rho'^{(2)}_1(R) = \Gamma^{(2)} \e^{-\kappa^{(2)} R/a} ,
\end{equation}
where $\Gamma^{(2)}$ and $\kappa^{(2)}$ are real coefficients. Upon identification of the two formulas, the normalized purity is shown to follow as
\begin{equation} \label{eq:p2_norm_ident}
    \bar p_2 = \frac{1}{K^2} \e^{-\kappa^{(2)} R_1 R_2/a^2} \left(\frac{\dim^4(J)\, \Gamma^{(2)}}{\rho^{(2)}_1}\right)^{(R_1+R_2)/a} .
\end{equation}
Notably, this structural form of $\bar p_2$ is very similar to that of the Wilson loop expectation value found in \cite{Zohar:2021wqy}. In particular, the dominant eigenvalue $\rho'^{(1)}_1$ of a single transfer row with two gauge field insertions was found to obey the same structural decay behavior as in \eqref{eq:rho1_form}, i.e.\ $\rho'^{(1)}_1 = \Gamma^{(1)} \e^{-\kappa^{(1)} R}$. Moreover, the distinction between the two possible cases thereof -- spatial independence ($\kappa^{(1)}=0$) and exponential decay ($\kappa^{(1)}>0$) -- leads to the Wilson loop perimeter law and Wilson loop area law, respectively. In the present context, we are instead interested in the behavior of the normalized R\'enyi entanglement entropy and want to explore the same corresponding regimes for $\kappa^{(2)}$. From the definition \eqref{eq:S_n_norm}, we have
\begin{equation} \label{eq:S2_PEPS_full}
    \bar S_2 = -\ln \bar p_2 = 2\ln K + \frac{\kappa^{(2)}R_1R_2}{a^2} - \frac{R_1+R_2}{a} \ln \left(\frac{\dim^4(J)\, \Gamma^{(2)}}{\rho^{(2)}_1}\right) .
\end{equation}
The first term in \eqref{eq:S2_PEPS_full} is simply a constant. The last term depends on the perimeter of the subsystem, which is multiplied with the logarithm of the ratio of dominant eigenvalues. As such, it is a manifestation of the entanglement area law. The second term, on the other hand, depends on the area of $A$ and hence would represent an entanglement volume law. From our physical expectations and the PEPS construction, we can exclude the existence of such a term: For a PEPS with bond dimension $\chi$ and a subsystem with perimeter $P$ (number of boundary lattice sites), one can easily show (using eqs.~\eqref{eq:S_n_reg} and \eqref{eq:S_1}) that $S_n \le S_1 \le P\ln\chi$. The PEPS ansatz naturally implements the entanglement area law. We therefore can conclude that the dominant eigenvalue of the transfer row $E^{(2)}_{||}$ may not have an exponential decay,
\begin{equation}
    \kappa^{(2)} \overset{!}{=} 0 ,
\end{equation}
i.e.\ $\rho'^{(2)}_1$ does not depend on the dimensions of the subregion, $\rho'^{(2)}_1 \ne \rho'^{(2)}_1(R)$. Using the notation $\rho'^{(2)}_1 \equiv \Gamma^{(2)} = \text{const}$, we write the final result for $\bar S_2$ as
\begin{equation} \label{eq:S2_PEPS_final}
    \bar S_2 = 2\ln K - \frac{R_1+R_2}{a} \ln \left(\frac{\dim^4(J)\, \rho'^{(2)}_1}{\rho^{(2)}_1}\right) .
\end{equation}

Formula \eqref{eq:S2_PEPS_final} is the first main result of this paper. It allows to calculate the second normalized R\'enyi entanglement entropy in the thermodynamic limit for an arbitrary 2$D$ LGT in terms of properties of transfer matrices. In contrast to the complementary results for the Wilson loop expectation value in \cite{Zohar:2021wqy}, which are also based on a transfer operator construction, we here cannot distinguish between two different geometric regimes. While the Wilson loop perimeter vs.\ area law detects the presence of a deconfined or confined phase, the R\'enyi entropy always exhibits an entanglement area law, and has no other apparent phase differences. On these general grounds, the R\'enyi entanglement entropy thus cannot serve as an equivalent probe of (de)confinement in pure LGTs. Nevertheless, as we will demonstrate in the next section, (de)confinement properties do leave an imprint on the behavior of entanglement. This, however, requires the study of parameter regimes for specific LGTs. 

Some further properties of $\bar S_2$ can be concluded from \eqref{eq:S2_PEPS_final}. First, since the entropy is manifestly positive, $\bar S_2 \ge 0$, we infer $\rho'^{(2)}_1 \le \rho^{(2)}_1$, which enforces a positive contribution through the logarithmic factor. 
The second important observation concerns the opposing infrared (IR) and ultraviolet (UV) limits. From our transfer operator derivation we see that an infinitely large subsystem ($R_{1,2} \to \infty$) causes an IR divergence in \eqref{eq:S2_PEPS_final}, while in the continuum limit $a \to 0$ an UV divergence emerges. Both are expected features of entanglement quantities in QFTs \cite{Headrick:2019eth}.

\subsubsection{Generalization to higher orders}

The transfer operator construction can be directly extended to calculate normalized R\'enyi entanglement entropies \eqref{eq:S_n_norm} of (integer) order $n>2$. For that purpose we consider the quantity
\begin{equation} \label{eq:p_n}
    p_n \equiv \Tr[\rho_A^n] ,
\end{equation}
which is the higher-order generalization of the purity \eqref{eq:p_2}. For a MPS, $p_n$ is structurally constructed in the same way as exemplified in \eqref{eq:p2_MPS} by adding the corresponding layers of the quantum state $\ket{\Psi}$ and its adjoint $\bra{\Psi}$. As for $p_2$, the resulting tensor network for $p_n$ is uniformly composed out of transfer operators and two insertions of boundary operators $X^{(n)}$. Repeating the same scheme as in \eqref{eq:p2_bdy} for the right boundary, one finds that $X^{(n)}$ acts as the following index map
\begin{equation} \label{eq:X_n}
    X^{(n)}: \{ 2i-1, 2i \} \mapsto \{ 2i-1, 2i+2 \} ,\qquad 1 \le i \le n .
\end{equation}
As in section \ref{sec:idea}, we have labelled the indices from top to bottom as $1,2,\ldots,2n$, and assumed periodicity for the last one, i.e.\ $\{2n\} \mapsto \{2\}$. At the left boundary, the mapping is inverted, which is described by the transposed operator $X^{(n)\mathsf{T}}$. By contracting the whole tensor network, $p_n$ is calculated as the trace.

We can directly generalize this construction to gauged PEPS by repeating the idea of lattice tiling. Defining the subsystem as before, and including the normalization factor, we then have
\begin{equation} \label{eq:pn_PEPS}
    \bar p_n \equiv \frac{\Tr[\rho_A^n]}{\Tr^n[\rho_A]} = \frac{\Tr\left[\mathcal X^{(n)\mathsf{T}}(R_1) E^{(n)R_2/a}_{||}(R_1) \mathcal X^{(n)}(R_1) E^{(n) N_2-R_2/a}\right]}{\Tr^n[E^{(1)N_1}]} .
\end{equation}
Here, the basic element is the plaquette operator
\begin{equation}
    \mathcal T^{(n)} \equiv \tau_0^{\otimes n} ,
\end{equation}
which contains $n$ tensor product copies of the transfer operator. Based on this construction, we have the following generalized definitions:
\begin{align}
\mathcal X^{(n)}(R_1) &\equiv X^{(n)\otimes R_1/a}\,\cdot\, \mathds{1}^{\otimes N_1-R_1/a} ,\nonumber\\
E_{||}^{(n)}(R_1) &= \Tr_{\text{row}}\left[X^{(n)\mathsf{T}} \otimes \mathcal T^{(n)\otimes R_1/a} \otimes X^{(n)} \otimes \mathcal T^{(n) \otimes N_1-R_1/a}\right] , \nonumber\\
E^{(n)} &= \Tr_{\text{row}}\left[\mathcal T^{(n)\otimes N_1} \right] . \nonumber
\end{align}
As in the previous section, we can make use of the eigenvalue decomposition and demand the invariance of the contraction scheme under the order. This directly leads to the formula
\begin{equation} \label{eq:Sn_PEPS_final}
    \bar S_n = \frac{1}{1-n} \ln \bar p_n = \frac{1}{1-n} \left[-n\ln K + \frac{(R_1+R_2)}{a} \ln\left(\frac{\dim^{2n}(J)\, \rho'^{(n)}_1}{\rho^{(n)}_1}\right)\right] .
\end{equation}
Expression \eqref{eq:Sn_PEPS_final} is valid in the thermodynamic limit for any LGT PEPS. It generalizes \eqref{eq:S2_PEPS_final} and is the second main result of this paper. As in the previous discussion, the dominant eigenvalue $\rho'^{(n)}_1$ of $E_{||}^{(n)}$ may not depend on the subsystem dimensions to forbid an entanglement volume law. For the dominant eigenvalue $\rho^{(n)}_1$ of $E^{(n)}$, the relation $\rho^{(n)}_1 = \rho^{(1)n}_1$ holds.

It becomes discernible that the previous conclusions for the entanglement behavior hold equally for arbitrary $\bar S_n$. That is, eq.~\eqref{eq:Sn_PEPS_final} manifests the entanglement area law for all normalized R\'enyi entanglement entropies. The results persist in the continuum limit, in which an UV divergence appears when $a \to 0$ is taken. 

These general results do not yet allow any conclusions regarding entanglement properties in the deconfined versus confined phase. In particular, we would expect larger amount of entanglement in the deconfined phase. The Wilson loop results in \cite{Zohar:2021wqy} shared that expectation through the observation that the eigenvectors of the transfer matrix $E^{(1)}$ are farther away from being product vectors in the deconfined phase. The result \eqref{eq:Sn_PEPS_final} would support that hypothesis, if the ratio of dominant eigenvalues, $\rho'^{(n)}_1 / \rho^{(n)}_1$, tends to decrease in the deconfined phase. However, from the general formalism outlined so far, we cannot conclude this behavior. It is therefore necessary to explicitly apply the transfer operator approach to a specific LGT and study its properties in the different phases.

As an alternative to the R\'enyi entanglement entropies, one could also compute the entropic c-functions \cite{Casini:2004bw,Casini:2006es,Nishioka:2006gr}, which extract subleading corrections to the area law term of the entropies themselves. This calculation, however, is only meaningful if the entangling surface is independent from the size of the subregion. Such a scenario exists for example when one considers an infinitely long slab as $A$ (with width $R_1$ and length $R_2 \to \infty$), which differs from our assumption in \eqref{eq:TD_limit}. The entropic c-function is then proportional to the derivative of $S_n$ w.r.t.\ $R_1$. This quantity is UV-finite and counts the degrees of freedom at the respective length (or energy) scale \cite{Nishioka:2006gr}. Such a setup has been implicitly employed in previous holographic explorations, and could also be used to study the limit of large-$N_c$ gauge theories in an extension of our present analysis. Interestingly, in the recent work \cite{Bulgarelli:2023ofi}, the c-function was calculated for the classical 3$D$ Ising model, which is dual to the $\mathds{Z}_2$ LGT that we study in the next section.

\section{Entanglement and (de)confinement for the $\mathds{Z}_2$ LGT}
\label{sec:Z2}

In this section we apply the transfer operator approach to the $\mathds{Z}_2$ LGT PEPS as an explicit example. By comparing with the Wilson loop results of \cite{Zohar:2021wqy}, we connect the entanglement characteristics to (de)confinement properties using numerical calculations.

\subsection{Transfer operator setup}

We consider a $\mathds{Z}_2$ LGT, for which the physical gauge fields are placed on the links of a 2$D$ lattice. The two-dimensional group Hilbert space on each link is spanned by the spin representations $j = +,-$. Under the action of the $x$-Pauli matrix as the Hermitian group element operator, spins are inverted, $\sigma^x \ket{\pm} = \ket{\mp}$. Local gauge invariance is given w.r.t.\ the operator 
\begin{equation}
 \hat\Theta(\bm{x}) = \sigma^z_{\bm{x},1} \sigma^z_{\bm{x},2} \sigma^z_{\bm{x}-\hat{\bm{e}}_1,1} \sigma^z_{\bm{x}-\hat{\bm{e}}_2,2} ,
\end{equation}
i.e.\ $\hat\Theta(\bm{x}) \ket{\psi} = \ket{\psi}$ $\forall \bm{x} \in \mathds{Z}^2$. Here, the $z$-Pauli matrix is the group operator acting as $\sigma^z \ket{\pm} = \pm \ket{\pm}$. 
The second group operation is represented by the trivial identity operator. In contrast to the most general form \eqref{eq:gauge_trafo_general} of the gauge transformation, we do not have any difference between left and right group operations here.

We consider the most general gauged PEPS ansatz, which is translationally and rotationally invariant. Moreover, we assume a minimal construction with the smallest possible bond dimension, i.e.\ physical and virtual spaces are identically spanned by the spin states $\ket{\pm}$. As shown in \cite{Zohar:2021wqy}, there are only 4 parameters necessary to fully encompass all nonvanishing gauged PEPS tensors $A^{st}_{ruld}$ ($s,t,r,u,l,d = \pm$) as follows:
\begin{align} \label{eq:fluxes_Z2}
    A^{++}_{++++} &\equiv \vcenter{\hbox{\includegraphics[width=0.06\columnwidth]{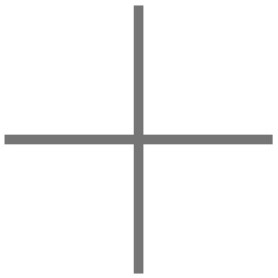}}} = \alpha, \nonumber\\
    A^{--}_{--++} &\equiv \vcenter{\hbox{\includegraphics[width=0.06\columnwidth]{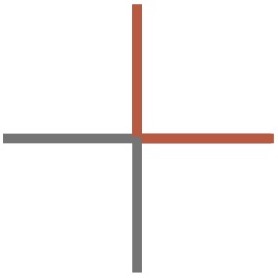}}} =
    A^{-+}_{-++-} \equiv \vcenter{\hbox{\includegraphics[width=0.06\columnwidth]{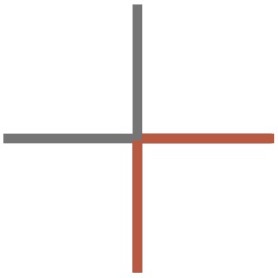}}} =
    A^{++}_{++--} \equiv \vcenter{\hbox{\includegraphics[width=0.06\columnwidth]{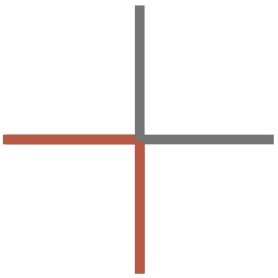}}} =
    A^{+-}_{+--+} \equiv \vcenter{\hbox{\includegraphics[width=0.06\columnwidth]{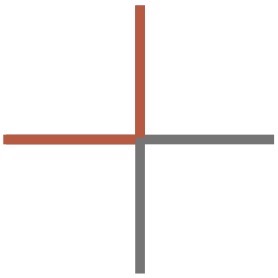}}} = \beta, \nonumber\\
    A^{-+}_{-+-+} &\equiv \vcenter{\hbox{\includegraphics[width=0.06\columnwidth]{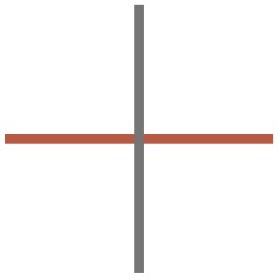}}} =
    A^{+-}_{+-+-} \equiv \vcenter{\hbox{\includegraphics[width=0.06\columnwidth]{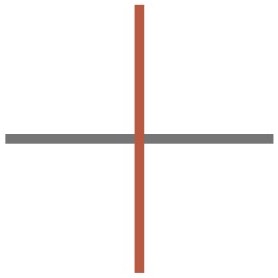}}} = \gamma, \nonumber\\
    A^{--}_{----} &\equiv \vcenter{\hbox{\includegraphics[width=0.06\columnwidth]{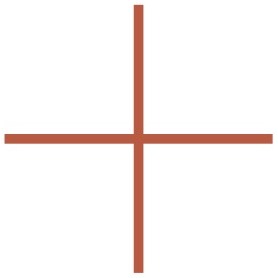}}} = \delta .
\end{align}
In this setup, we choose to interpret $\ket{+}$ as a flux-free state, and $\ket{-}$ as the flux-carrying one. Hence, the parameters $\alpha,\ldots,\delta$ in \eqref{eq:fluxes_Z2} correspond to zero flux, corner fluxes, straight line fluxes and crossing flux lines.\,\footnote{Note that in \eqref{eq:fluxes_Z2} only virtual indices are graphically shown. The physical fluxes follow from the gauge symmetry constraints as $j_s \sim j_r$ and $j_t \sim j_u$. Any flux $j=-$ is represented by a red line, and zero fluxes $j=+$ by gray lines.}

The single-site transfer operator $T^{(1)}$ is then constructed in the same way as in \cite{Zohar:2021wqy} using \eqref{eq:T1} and \eqref{eq:T1_elem}. In fact, we can directly infer the operator $\hat\tau_0$, which takes the simplifying gauge symmetry constraints into account when calculating traces in the 2$D$ lattice, by identifying the vector space elements of the transfer operator legs. There are two on-leg singlets, $\ket{0(+)} \equiv \ket{++} = \ket{+}\otimes\ket{s}$ and $\ket{0(-) \equiv \ket{--}} = \ket{-}\otimes\ket{s}$, which factorize the on-leg Hilbert space into a product of two spin spaces in which the first one is either a $\ket{+}$ or $\ket{-}$ state, and the second one detects a singlet $\ket{s}$. 
When choosing the index ordering $\{ \ket{0(+)}\bra{0(+)}, \ket{0(-)}\bra{0(-)}, \ket{0(+)}\bra{0(-)}, \ket{0(-)}\bra{0(+)} \}$, the transfer operator takes the explicit form
\begin{equation} \label{eq:tau0_Z2}
	\hat{\tau}_0 = \begin{pmatrix}
    |\alpha|^2 & |\gamma|^2 & 0 & 0 \\
    |\gamma|^2 & |\delta|^2 & 0 & 0 \\
    0 & 0 & |\beta|^2 & |\beta|^2 \\
    0 & 0 & |\beta|^2 & |\beta|^2
    \end{pmatrix} ,
\end{equation}
where in this convention the rows correspond to the horizontal PEPS directions ($l,r$), while the columns correspond to the vertical ones ($d,u$).\footnote{As elaborated in \cite{Zohar:2021wqy}, the upper block only depends on the amplitudes for which the flux does not change direction, i.e.\ representations are not changed horizontally or vertically. The lower block is associated with corner fluxes for which the representation changes.}
By diagonalizing \eqref{eq:tau0_Z2}, we can rewrite $\tau_0$ in the form \eqref{eq:tau0_VLV}. The matrix $V$ takes the form
\begin{equation} \label{eq:V_Z2}
	V = \begin{pmatrix}
    v_{11}(\alpha,\gamma,\delta) & v_{12}(\alpha,\gamma,\delta) & 0 & 0 \\
    v_{21}(\alpha,\gamma,\delta) & v_{22}(\alpha,\gamma,\delta) & 0 & 0 \\
    0 & 0 & \frac{1}{\sqrt{2}} & -\frac{1}{\sqrt{2}} \\
    0 & 0 & \frac{1}{\sqrt{2}} & \frac{1}{\sqrt{2}}
    \end{pmatrix} ,
\end{equation}
where the entries $v_{11}, v_{12}, v_{21}, v_{22}$ are determined numerically for chosen values of the PEPS parameters.\footnote{Note that the row indices in \eqref{eq:V_Z2} are the same as for $\hat\tau_0$, while the column ones correspond to $\mu = 1,\ldots,4$.}
With the eigenvalues
\begin{equation}
    \lambda_{1,2} = \frac{1}{2}\left( |\alpha|^2 + |\delta|^2 \pm \sqrt{\left(|\alpha|^2-|\delta|^2\right)^2+4|\gamma|^4} \right) ,\quad \lambda_3 = 2|\beta|^2 ,\quad \lambda_4 = 0 ,
\end{equation}
and the operators
\begin{align}
    \hat M_1 &= v_{11}\begin{pmatrix}1&0\\0&0\end{pmatrix} + v_{21} \begin{pmatrix}0&0\\0&1\end{pmatrix} ,\quad&
    \hat M_2 &= v_{12}\begin{pmatrix}1&0\\0&0\end{pmatrix} + v_{22} \begin{pmatrix}0&0\\0&1\end{pmatrix} \nonumber\\
    \hat M_3 &= \frac{1}{\sqrt{2}}\sigma^x ,\quad&
    \hat M_4 &= -\frac{i}{\sqrt{2}} ,\nonumber
\end{align}
the transfer operator then takes the desired form \eqref{eq:tau0_Mform}.\footnote{The subsequent numerical analyses can be made more efficient by neglecting $\hat M_4$, since its corresponding eigenvalue vanishes. The singlet contributions $\ket{s}\bra{s}$ to the operators $\hat M_\mu$ are neglected because they multiply everything equally.}

As outlined in the previous section, when calculating purities and R\'enyi entanglement entropies, we need to construct transfer rows. For the following studies, we restrict ourselves to the second order ($n=2$), and calculate the normalized R\'enyi entanglement entropy $\bar S_2$. Using the above setup, we can construct the operator $E^{(1)}$ directly as
\begin{equation}
E^{(1)} = \Tr_{\text{row}}[\tau_0^{\otimes N_1}] = \sum_{\{\mu\}} (\lambda_{\mu_1} \cdot\ldots\cdot \lambda_{\mu_{N_1}}) \Tr[M_{\mu_1} \cdot\ldots\cdot M_{\mu_{N_1}}] M_{\mu_1} \otimes\ldots\otimes M_{\mu_{N_1}} ,
\end{equation}
where the indices $\mu_x = 1,2,3,4$ are taken for each lattice site $x=1,\ldots,N_1 $ along a row. The operator $E^{(2)}=E^{(1)} \otimes E^{(1)}$ follows directly as a tensor product copy of $E^{(1)}$. However, for numerically evaluating this expression in matrix form, it is more convenient to choose the tensor product structure site-wise. We therefore have
\begin{align}
E^{(2)} = \sum_{\{\mu\},\{\nu\}} &(\lambda_{\mu_1} \cdot\ldots\cdot \lambda_{\mu_{N_1}}) (\lambda_{\nu_1} \cdot\ldots\cdot \lambda_{\nu_{N_1}}) \Tr[M_{\mu_1} \cdot\ldots\cdot M_{\mu_{N_1}}] \Tr[M_{\nu_1} \cdot\ldots\cdot M_{\nu_{N_1}}] \nonumber \\
&(M_{\mu_1} \otimes M_{\nu_1}) \otimes\ldots\otimes (M_{\mu_{N_1}} \otimes M_{\nu_{N_1}}) ,
\end{align}
where now the set of both $\{\mu\}$ and $\{\nu\}$ indices run over all values independently.
For the transfer row $E^{(2)}_{||}$, the single-site boundary operator $X$, given as
\begin{equation}
X = \begin{pmatrix}
	1&0&0&0\\
	0&0&0&1\\
	0&0&1&0\\
	0&1&0&0 
\end{pmatrix} ,
\end{equation}
is additionally inserted at positions $x=1$ and $x=R_1$ as follows
\begin{align}
E^{(2)}_{||}(R_1) =& \sum_{\{\mu\},\{\nu\}} (\lambda_{\mu_1} \cdot\ldots\cdot \lambda_{\mu_{N_1}}) (\lambda_{\nu_1} \cdot\ldots\cdot \lambda_{\nu_{N_1}}) 
\Tr\Bigl[ X (M_{\mu_1} \otimes M_{\nu_1}) \cdot\ldots\cdot (M_{\mu_{R_1}} \otimes M_{\nu_{R_1}}) X \cdot \nonumber\\
&(M_{\mu_{R_1+1}} \otimes M_{\nu_{R_1+1}}) \cdot\ldots\cdot (M_{\mu_{N_1}} \otimes M_{\nu_{N_1}}) \Bigr] (M_{\mu_1} \otimes M_{\nu_1}) \otimes\ldots\otimes (M_{\mu_{N_1}} \otimes M_{\nu_{N_1}}) .
\end{align}
Defining the boundary row $\mathcal X(R_1)$ as
\begin{equation}
\mathcal X(R_1) \equiv X(x=1) \otimes \ldots \otimes X(x=R_1) \otimes \mathds{1}_{(4)}(x=R_1+1) \otimes \ldots \otimes \mathds{1}_{(4)}(x=N_1) ,
\end{equation}
we can numerically calculate the normalized purity as follows
\begin{equation}
\bar p_2 = \frac{\Tr[\mathcal X(R_1) E^{(2)R_2}_{||}(R_1) \mathcal X(R_1) E^{(2) N_2-R_2}]}{\Tr^2[E^{(1) N_2}]} .
\end{equation}
The normalized R\'enyi entanglement entropy is finally given by
\begin{equation}
\bar S_2 = -\ln \bar p_2 .
\end{equation}

\subsection{Numerical results}

In full generality, the outlined $\mathds{Z}_2$ LGT PEPS setup has a 4-dimensional complex parameter space $\{\alpha,\beta,\gamma,\delta\} \in \mathds{C}^4$. The analytical and numerical Wilson loop studies in \cite{Zohar:2021wqy} revealed that it is in fact a nontrivial task to identify the confined phase in this theory through a Wilson loop area law. In particular, it was shown that only in the perturbative regime $|\beta| \ll |\delta| \lesssim |\alpha|$ and $\gamma=0$, one finds a Wilson loop area law, i.e.\ a confining phase, as long as $|\delta| \ne |\alpha|$. In the case $|\delta| = |\alpha|$, there exists a deconfined phase. 

In the first part of our numerical studies, we want to confirm the existence of an entanglement area law as found in our general analytical treatment. In all subsequent analyses, we restrict ourselves to the real and positive parameter space $\mathds{R}_+^4$. We consider the confining phase in the perturbative regime and define a thin but long 2$D$ lattice of dimensions $N_1=4$, $N_2=100$, and set $\alpha=1, \beta=0.1, \gamma=0, \delta=0.95$. We implicitly assume unit lattice spacing $a=1$.

\begin{figure}[t]
    \centering
   \includegraphics[width=0.49\textwidth]{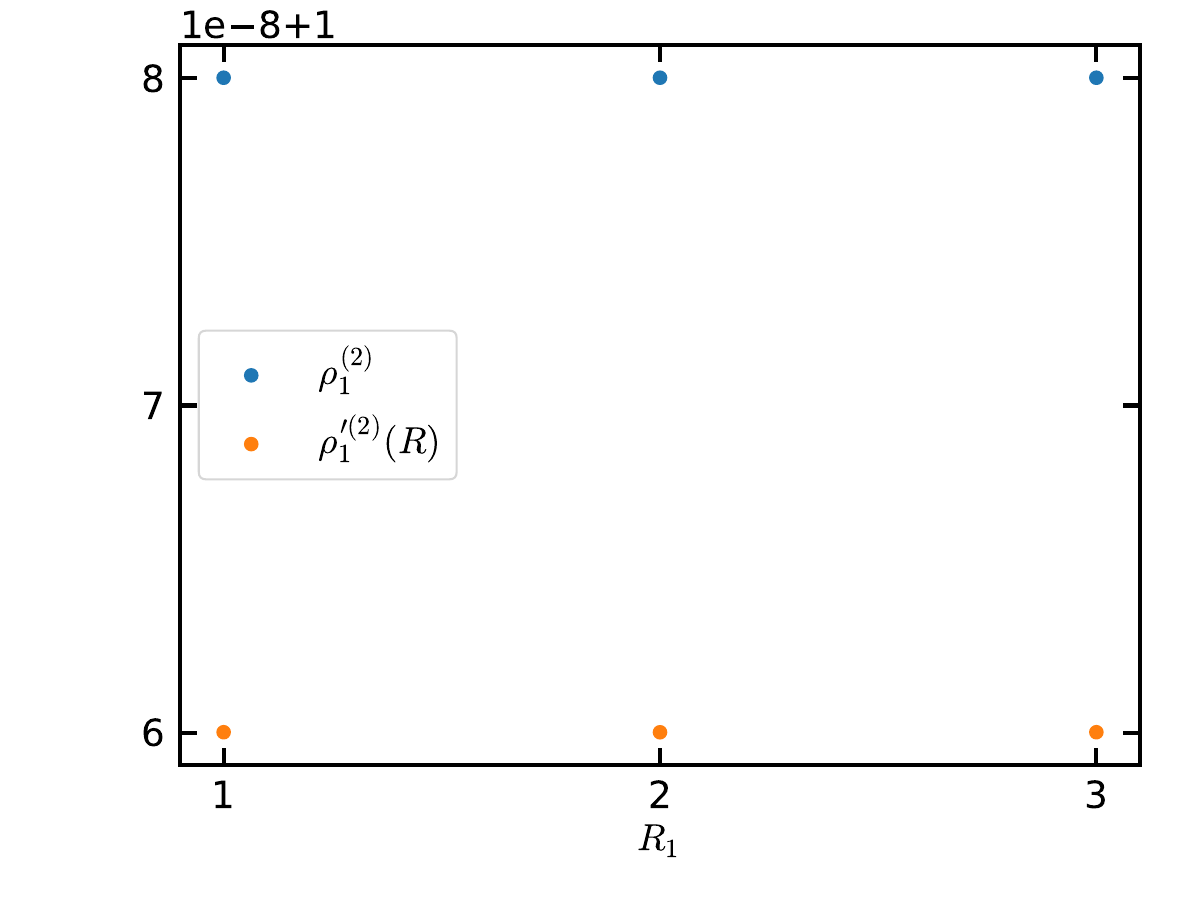}
   \includegraphics[width=0.49\textwidth]{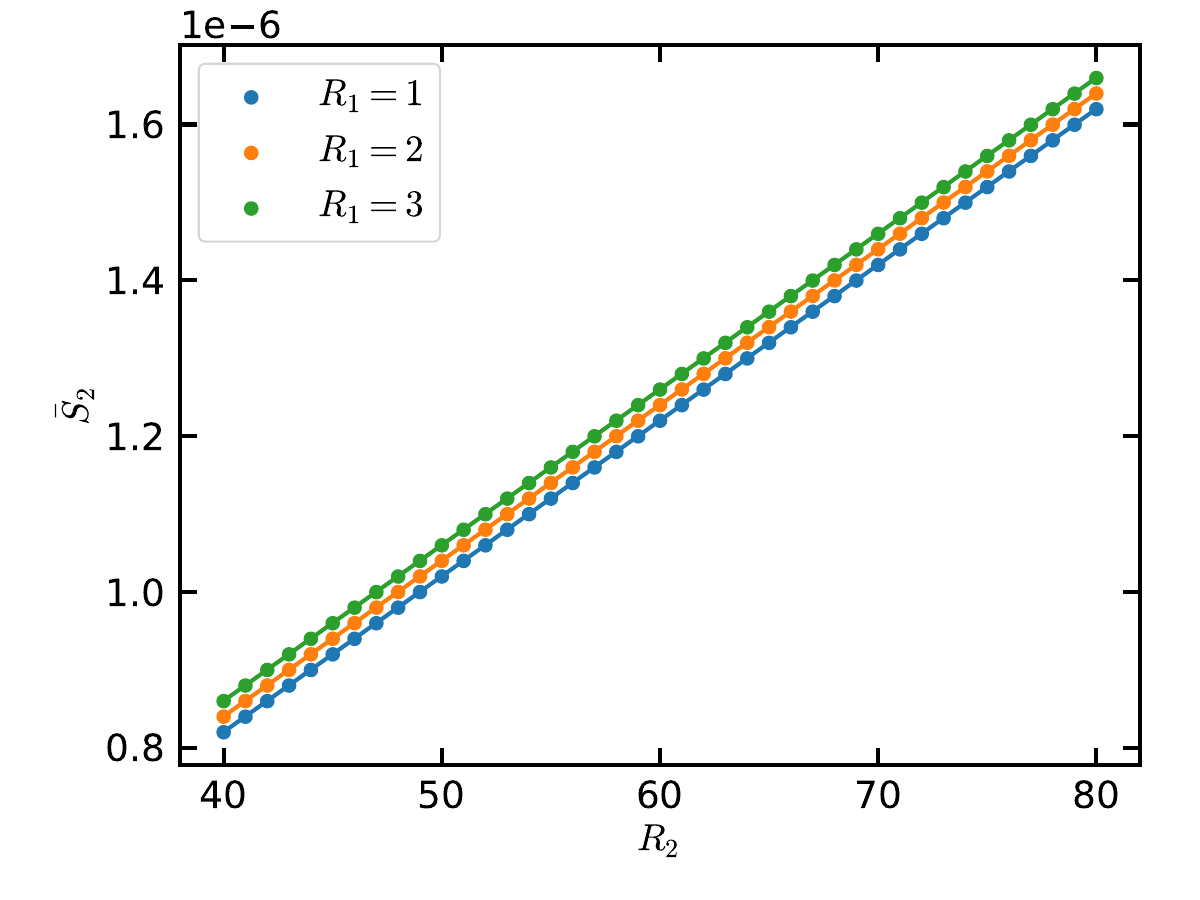}
    \caption{Left: Independence of the dominant transfer operator eigenvalues $\rho_1^{(2)}$ and $\rho_1^{\prime(2)}$ from the subsystem width $R_1$. 
    Right: Dependence of the normalized R\'enyi entanglement entropy $\bar S_2$ on the subsystem length $R_2$ for several values $R_1$ (colored dots). The colored curves represent the linear prediction in the thermodynamic limit for each case, confirming the entanglement area law of the numerical data. See text for further explanations.}
    \label{fig:s2_vs_R2_pert}
\end{figure}
The left panel in Fig.\,\ref{fig:s2_vs_R2_pert} shows the dominant eigenvalue $\rho_1^{\prime(2)}$ of $E^{(2)}_{||}(R_1)$ (orange dots) for three different subsystem widths $R_1=1,2,3$. The numerical values are identical for all cases and hence confirm our theoretical prediction that $\rho_1^{\prime(2)}$ is independent of $R_1$. As a comparison, we also plot the dominant eigenvalue $\rho_1^{(2)}$ of $E^{(2)}$ (blue dots), which allows for a positive contribution to $\bar S_2$ in the analytical formula \eqref{eq:S2_PEPS_final} due to $\rho_1^{\prime(2)} < \rho_1^{(2)}$. 

The right panel in Fig.\,\ref{fig:s2_vs_R2_pert} shows the resulting values of $\bar S_2$ for a large range of subsystem lengths $R_2$ and several parameter values $R_1$ (colored dots). On top of the data points, linear functions are shown for each parameter, which follow directly from the analytical result \eqref{eq:S2_PEPS_final} when the previously determined numerical values of $\rho_1^{(2)}$ and $\rho_1^{\prime(2)}$ are inserted. The numerical data points agree perfectly with the linear prediction. These results confirm the presence of an entanglement area law. Moreover, it demonstrates that the chosen numerical setup already reproduces results in the thermodynamic limit, i.e.\ for infinitely large systems. These conclusions are independently also valid in other parameter ranges of the $\mathds{Z}_2$ LGT PEPS when a nonperturbative or deconfined regime is considered.

\begin{figure}[t]
    \centering
   \includegraphics[width=0.49\textwidth]{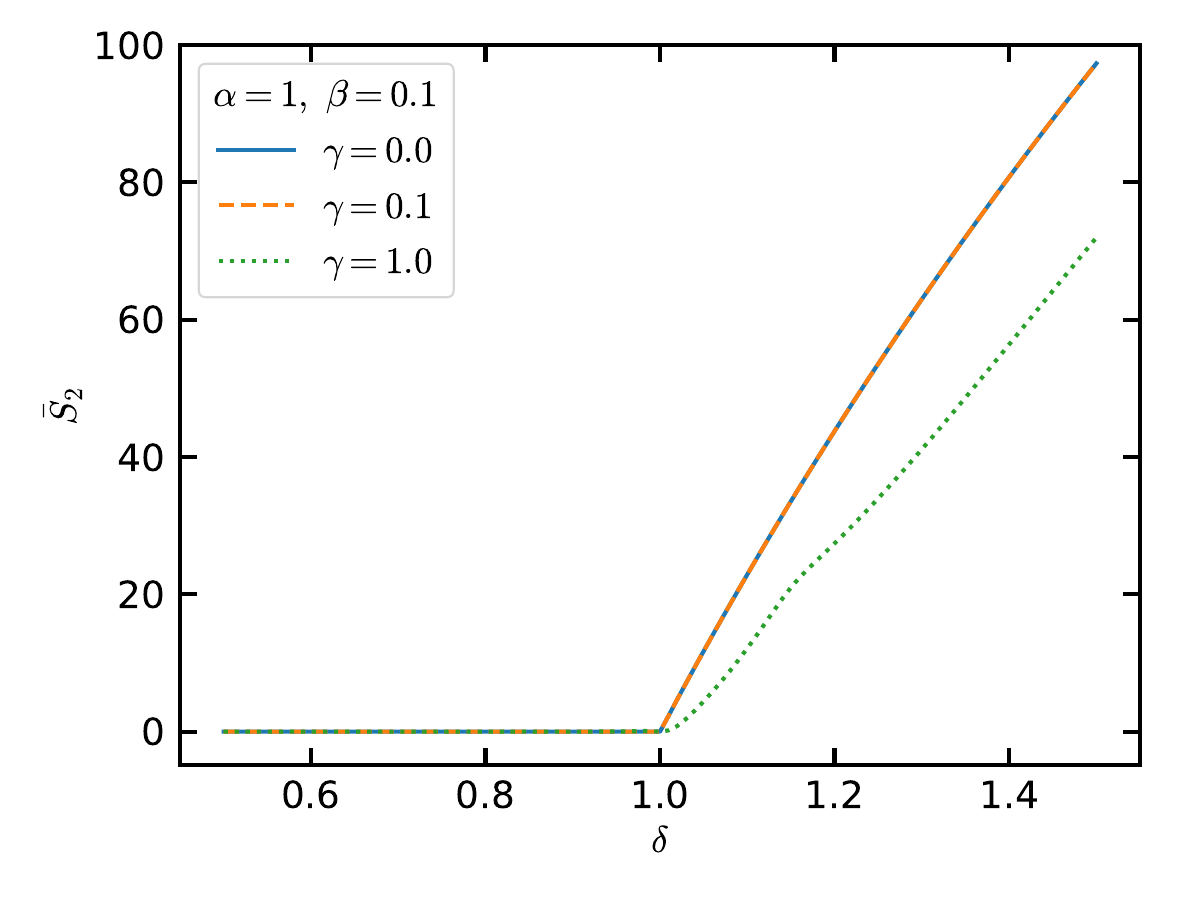}
    \caption{Dependence of $\bar S_2$ on the parameter $\delta$ in the perturbative regime. The data demonstrate that (de)confinement leaves an imprint on entanglement properties: At $\gamma=0$ (blue curve), the theory is confined except at the singular point $\alpha = \delta = 1$ as signaled by a nonanalytic kink. The kink disappears towards larger values of $\gamma$ (orange and green curve) when the theory becomes deconfined in the whole parameter space. See text for further explanations.}
    \label{fig:s2_vs_delta_pert}
\end{figure}
Our analytical studies in the previous section have shown, in contrast to the Wilson loop expectation value, that there is no geometric distinction between the confined and deconfined phase for the R\'enyi entanglement entropies in 2$D$ LGT PEPS. To study the interplay of (de)confinement and entanglement in this situation, it is instead necessary to consider the dependence on the underlying parameter space of the theory. For that purpose we study in Fig.\,\ref{fig:s2_vs_delta_pert} the dependence of $\bar S_2$ on $\delta$ for $\alpha=1$ and $\beta=0.1$, i.e.\ in the perturbative regime as before. At $\gamma=0$ (blue curve), $\bar S_2$ is monotonously increasing on a very small scale as long as $\delta < 1$, while it is rapidly increasing for $\delta >1$. At $\delta = 1$, there is a nonanalytic kink in the data. This point $\alpha=\delta$ agrees precisely with the singular deconfined point in this regime. In other words, deconfinement leaves a clear imprint on $\bar S_2$. However, it does not match the expectation that the deconfined phase should be associated with a larger value of $\bar S_2$ since the amount of entanglement is expected to increase. 

One important result of the Wilson loop study in \cite{Zohar:2021wqy} was the following conclusion: In the confining phase at $\gamma=0$ of the perturbative regime, the eigenvectors of the transfer row $E^{(1)}$ are product vectors. If $\gamma$ is switched on, the eigenvectors are taken further away from product form; the Wilson loop area law is broken and the phase becomes deconfined. In Fig.\,\ref{fig:s2_vs_delta_pert} we study the corresponding behavior of the R\'enyi entanglement entropies for such a scenario. The orange and green curve show the behavior of $\bar S_2$ for $\gamma = 0.1$ and $\gamma=1$, respectively. While the $\gamma = 0.1$ curve is nearly identical to the previously discussed one at $\gamma=0$, $\bar S_2$ deviates for the largest plotted value of $\gamma$. Specifically, the nonanalyticity at $\delta=1$ disappears in the latter case and the values are lower in the regime $\delta > 1$.

\begin{figure}[t]
    \centering
   \includegraphics[width=\textwidth]{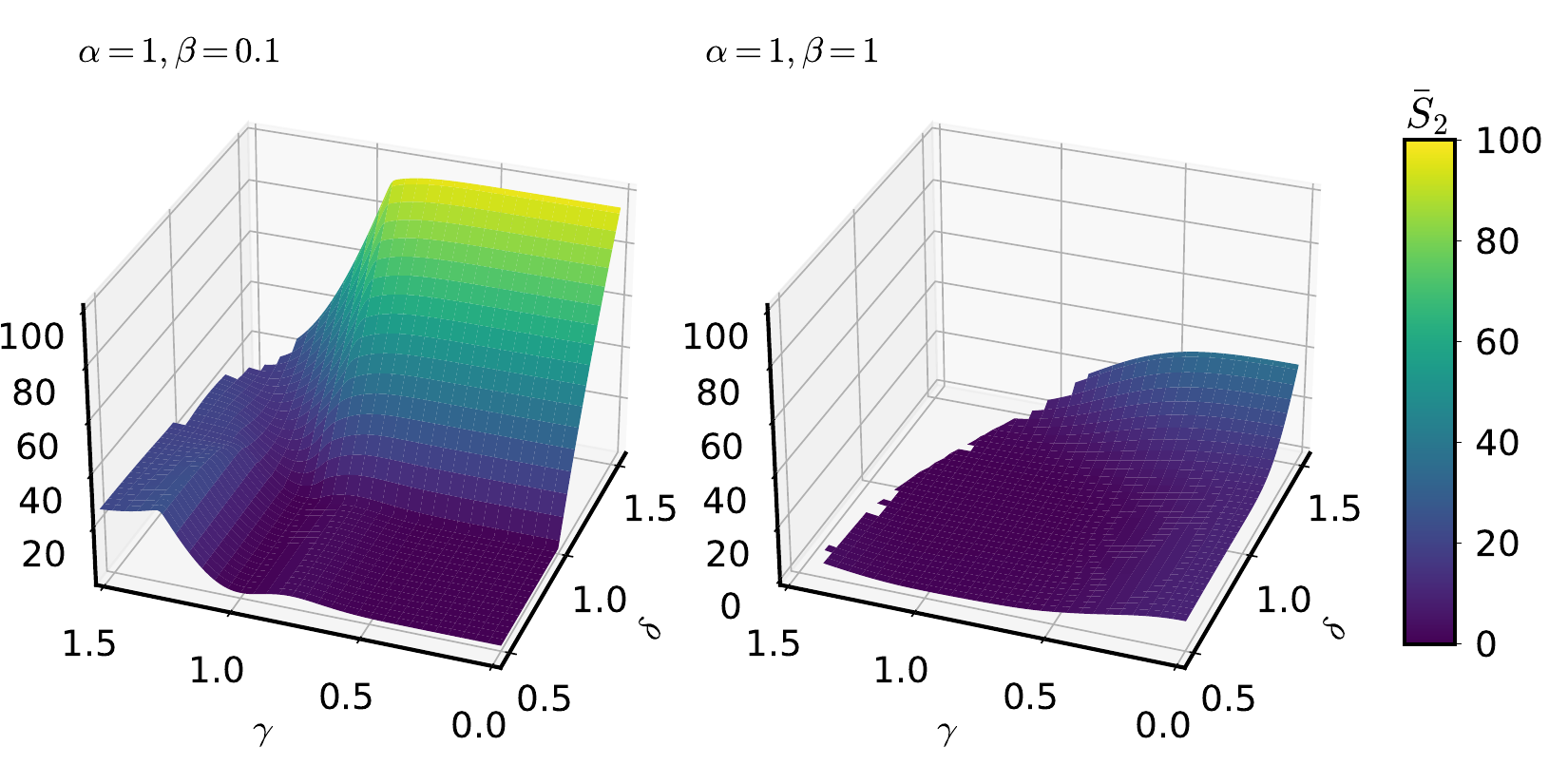}
    \caption{Behavior of $\bar S_2$ in the $\gamma-\delta$ parameter plane. The left plot is in the perturbative regime, the right one in the nonperturbative regime. See text for detailed discussions.}
    \label{fig:s2_vs_delta_gamma_both}
\end{figure}
In Fig.\,\ref{fig:s2_vs_delta_gamma_both} we extend these analyses by showing $\bar S_2$ in the whole $\gamma-\delta$ parameter plane. The left plot is in the perturbative regime at $\beta = 0.1$, and the right one in the nonperturbative regime at $\beta = 1$. For the perturbative case it becomes discernible that there is a nontrivial structure with local minima and maxima as $\gamma$ is increasing (i.e.\ when the theory becomes deconfined). The previously observed rapid increase of $\bar S_2$ towards large values of $\delta$ persists only as long as $\gamma \lesssim 1$. In the nonperturbative regime, i.e.\ for a completely deconfined phase, $\bar S_2$ is instead monotonously decreasing towards larger values of $\gamma$.

The conclusions of these numerical analyses are the following. First, our transfer operator approach allows us to reliably calculate R\'enyi entanglement entropies for arbitrary LGTs in two dimensions, which we demonstrated for the example of $\mathds{Z}_2$. In fact, we could achieve results for infinitely large systems in the thermodynamic limit already for moderate system sizes. Second, when studying the interplay of (de)confinement and entanglement, it is necessary to derive and analyze the PEPS parameter space of the considered LGT. We then found that (de)confinement does leave an imprint on the behavior of the R\'enyi entanglement entropy for specific parameter dependencies. However, it becomes also clear that these entropies cannot serve as an equally reliably order parameter of (de)confinement as the Wilson loop expectation value. In particular, it is not discernible that $\bar S_2$ (or any higher order entropy) explicitly counts the number of degrees of freedom in our specific setup, which is in contrast to higher-dimensional results for the entanglement entropy $S_1$ (cf.\ the discussions in the introduction).

\section{Discussion and outlook}
\label{sec:summary}

Motivated by recent explorations in quantum information science, particle physics and holography, we have studied the interplay of entanglement and confinement for 2$D$ pure LGTs in this paper. We have developed a transfer operator approach based on PEPS, which allowed us to calculate normalized R\'enyi entanglement entropies \eqref{eq:S_n_norm} within an extended Hilbert space formalism. The basis of our approach is the observation that the relevant tensor network diagram is uniformly composed out of transfer operators and additional operator insertions along the subsystem boundary. The whole lattice therefore can be tiled by plaquettes of transfer and boundary operators. Using the gauge invariance of the full pure state density matrix, we then have analyzed modifications of the Gauss laws for the reduced density matrix along the subsystem boundary for this approach. 

Local properties of transfer operators on-site and along rows of the lattice turn out to dictate the long-range properties of the entanglement entropies. This is encapsulated in our final result \eqref{eq:Sn_PEPS_final}, which is valid in the thermodynamic limit. It exemplifies the entanglement area law and exhibits an UV divergence in the continuum limit. The transfer operator approach hence provides a powerful tool to study entanglement properties in arbitrary pure LGTs. However, in contrast to the Wilson loop expectation value as another nonlocal observable, the normalized R\'enyi entanglement entropies do not have different geometric regimes or other features indicating the presence of a confined or deconfined phase. 

We have explicitly applied our methodology to the $\mathds{Z}_2$ LGT PEPS with minimal bond dimension. Using numerical evaluations, we reproduced results for $\bar S_2$ in the thermodynamic limit already for moderate lattice dimensions. By comparing with Wilson loop results of \cite{Zohar:2021wqy}, we found that (de)confinement does leave imprints on the behavior of the R\'enyi entanglement entropy in specific parameter regimes. However, the entropies are not an unambiguous probe of confinement or deconfinement in the whole parameter space of the theory. 

For future studies, it would be highly interesting to include dynamical matter (i.e.\ fermionic degrees of freedom) to the setup. In this for the standard model relevant physical situation, the structure of the tensors and transfer matrices would be modified.
Recent results based on a gauged Gaussian fermionic PEPS ansatz \cite{Emonts:2022yom}, which employs sign-problem free Monte-Carlo techniques for tensor contractions \cite{Zohar:2017yxl}, could provide a promising starting point for that purpose. Moreover, the first explorations of this ansatz in 3+1$D$ \cite{Emonts:2023ttz} could be used to generalize our approach to higher dimensions.

\acknowledgments

We deeply appreciate the long-term collaboration with Noa Feldman and Moshe Goldstein, which fruitfully influenced this project and the companion paper \cite{resolved_project}.  
Discussions with Daniel Gonz\'alez-Cuadra, Yannick Meurice and Michael Smolkin are also greatfully acknowledged.
We are grateful to the long term workshop YITP-T-23-01 held at YITP, Kyoto University, where a part of this work was done.
JK is supported by the Israel Academy of Sciences and Humanities \& Council for Higher Education Excellence Fellowship Program for International Postdoctoral Researchers.


\appendix

\section{Spatial dependence of the dominant transfer matrix eigenvalue $\rho'^{(2)}_1$}
\label{app:proof}

In section~\ref{sec:thermolim} we have derived an analytical result for the normalized purity $\bar p_2$ in the thermodynamic limit. Due to the invariance of the contraction scheme, the result \eqref{eq:p2bar_PEPS_row}, following from a row-wise contraction, and \eqref{eq:p2bar_PEPS_column}, resulting from a column-wise tiling, have to be identical. For notational convenience we here use a $x-y$ coordinate system and define 
\begin{align} \label{eq:def1}
    f(x) &\equiv \left(\frac{\rho'^{(2)}_1(x)}{\rho^{(2)}_1}\right) ,\\
    g(x) &\equiv \sum_{i=1}^{K} \sum_{j=1}^{K'} \bra{w^{(2)}_i}\mathcal{X}(x)\ket{v'^{(2)}_j(x)} \bra{w'^{(2)}_j(x)}\mathcal{X}(x)\ket{v^{(2)}_i} .
\end{align}
Demanding the equivalence of both contraction schemes means
\begin{equation}
    f^y(x) g(x) \overset{!}{=} f^x(y) g(y) .
\end{equation}
Taking the natural logarithm of both sides gives
\begin{equation}
    y\ln[f(x)] + \ln[g(x)] = x\ln[f(y)] + \ln[g(y)] ,
\end{equation}
which, by dividing with $xy$, can be transformed into
\begin{equation} \label{eq:rel1}
    \frac{\ln[f(x)]}{x} + \frac{\ln[g(x)]}{xy} = \frac{\ln[f(y)]}{y} + \frac{\ln[g(y)]}{xy} .
\end{equation}
Defining 
\begin{equation} \label{eq:rel2}
    F(x) \equiv \frac{\ln[f(x)]}{x} ,\quad G(x) \equiv \frac{\ln[g(x)]}{x} ,
\end{equation}
and taking a partial derivative (denoted by a prime) of the previous relation \eqref{eq:rel1} w.r.t.\ $x$, we get
\begin{align}
    F'(x) + \frac{1}{y} G'(x) &= -\frac{1}{x^2} G(y) \\
    \Leftrightarrow\qquad  G(y) &= -x^2 F'(x) -\frac{x^2}{y} G'(x) .\label{eq:G_y}
\end{align}
Taking now also a derivative w.r.t.\ $y$ yields
\begin{align}
    G'(y) &= \frac{x^2}{y^2} G'(x) \\
    \Leftrightarrow\qquad x^2 G'(x) &= y^2 G'(y) \overset{!}{=} -c_1 .
\end{align}
In the last relation, we have achieved a separation of variables. For the relation to hold, both sides have to be identical to a constant, which we denote as $-c_1$. Integrating back gives us
\begin{equation}
    G(x) = \frac{c_1}{x} + c_2 ,
\end{equation}
where $c_2$ is some integration constant. Substituting the result into \eqref{eq:G_y} and integrating it allows us to deduce $F(x)$:
\begin{align}
    \frac{c_1}{y} + c_2 &= -x^2 F'(x) + \frac{c_1}{y} \\
    \Leftrightarrow\qquad F'(x) &= -\frac{c_2}{x^2} \\
    \Rightarrow\qquad F(x) &= \frac{c_2}{x} + c_3 .
\end{align}
Using the definitions \eqref{eq:rel2}, we can transform back to our original functions:
\begin{equation} \label{eq:rel3}
    f(x) = \e^{c_2+c_3x} ,\quad g(x) = \e^{c_1+c_2x} .
\end{equation}
Since $\rho_1^{(2)}$ in \eqref{eq:def1} is simply a constant, the solution for $f(x)$ implies that the dominant eigenvalue $\rho'^{(2)}_1$ must depend exponentially on the spatial coordinate $x$ (or $y$, depending on the contraction scheme),
\begin{equation}
    \rho'^{(2)}_1(x) = c_2 \e^{-c_3x} ,
\end{equation}
where we have renamed the arbitrary constants $c_2, c_3$. Moreover, the solutions \eqref{eq:rel3} imply the following general behavior of the normalized purity
\begin{equation}
    \bar p_2 \sim f^y(x) g(x) = \e^{c_2y+c_3xy}\e^{c_1+c_2x} = \e^{c_1+c_2(x+y)+c_3xy} .
\end{equation}
When taking the prefactor and conventions of section \ref{sec:thermolim} into account, the result can be written in the form \eqref{eq:p2_norm_ident}. The general argument in this proof holds equally for the structural form of the Wilson loop expectation value.

\bibliographystyle{JHEP}
\bibliography{biblio.bib}

\end{document}